\documentclass[revtex4]{emulateapj}
\usepackage{graphicx}
\usepackage{natbib}
\usepackage{longtable}
\usepackage{amsmath}
\usepackage{subfigure}
\usepackage{placeins}
\begin{document}
\title{Spectroscopic Binaries in the Orion Nebula Cluster and NGC 2264}
\author{Marina Kounkel\altaffilmark{1}, Lee Hartmann\altaffilmark{1}, John J. Tobin\altaffilmark{2}, Mario Mateo\altaffilmark{1}, John I. Bailey, III.\altaffilmark{1}, Meghin Spencer\altaffilmark{1}}
\altaffiltext{1}{Department of Astronomy, University of Michigan, 1085 S. University st., Ann Arbor, MI
48109, USA}
\altaffiltext{2}{Leiden Observatory, Leiden University, P.O. Box 9513, 2300-RA Leiden, The Netherlands}
\email{mkounkel@umich.edu}

\clearpage

\begin{abstract}
We examine the spectroscopic binary population for two massive nearby regions of clustered star formation, the Orion Nebula Cluster and NGC 2264, supplementing the data presented by Tobin et al. (2009, 2015) with more recent observations and more extensive analysis. The inferred multiplicity fraction up to 10 AU based on these observations  is $5.3\pm 1.2$\% for NGC 2264 and $5.8\pm 1.1$\% for the ONC; they are consistent with the distribution of binaries in the field in the relevant parameter range. Eight of the multiple systems in the sample have enough epochs to make an initial fit for the orbital parameters. Two of these sources are double-lined spectroscopic binaries; for them we determine the mass ratio.  Our reanalysis of the distribution of stellar radial velocities towards these clusters presents a significantly better agreement between stellar and gas kinematics than was previously thought. 
\end{abstract}

\clearpage

\section{Introduction}\label{sec:intro}

Approximately a half of Sun-like stars belong to binary or higher order multiple systems \citep{2010raghavan}, and this fraction increases for more massive primaries \citep[e.g.][]{1991duquennoy,2007Kouwenhoven,2013Rizzuto}. Observations of several nearby non-clustered star-forming regions and associations revealed that they contain a larger fraction of the multiple systems as compared with the significantly more evolved field stars on the other hand, high-resolution imaging suggests that some young clusters are deficient in binaries \citep[and references therein]{2013duchene}. Since most stars are thought to form in dense clusters \citep{2010adams}, studies of binary frequencies as a function of young cluster structure and dynamics can shed light on the processes behind the present day stellar multiplicity.

The Orion Nebula Cloud (ONC) and NGC 2264 are two of the closest regions of clustered star formation \citep[$d \sim$400 pc and $\sim900$ pc, ages 1---2 Myr and 1.5---3 Myr respectively,][]{2007menten, 2007sandstrom,1997sung,2009baxter, 1997Hillenbrand}. While the binarity of nearby stars is relatively well understood, it becomes increasingly more difficult to characterize the full membership of multiple stars in the young star forming regions due to their larger distances. Nonetheless, extensive studies of optical binaries in the ONC have led to interesting findings. Most of the studies concentrated on wider binaries that can be detected with photometric surveys, with smallest separations between two companions of $\sim$60 AU, set by the diffraction limit of the optics \citep{2006kohler,2007reipurth}. A recent survey of Class I and II stars in the Orion Molecular Clouds by \cite{hstbinary} revealed, that contrary to expectations, the densely populated regions have a larger fraction of the wide multiple systems than the diffusely populated ones, highlighting the need to reexamine the environmental dependence on the evolution of the multiplicity. Constructing a better defined sample of the close binary systems that can be obtained through multi-epoch spectroscopic monitoring is an important step in this process.

In addition, identifying spectroscopic binaries and removing them from the sample in the kinematic studies of star forming regions can help refine tests of cluster formation. There is debate over whether clusters form in a slow process, taking place in clouds initially supported by supersonic turbulence  \citep[e.g.][]{2006tan, 2012hennebelle}, or whether clusters form rapidly on the scale of a free-fall due to a gravitational collapse \citep[e.g.][]{2007hartmann, 2007elmegreen,2015kuznetsova}. Simulations have shown that in the case of the former, any subclustering in the initial environment would not change significantly in the cluster evolution; for the latter any substructure would rapidly dissipate in only a few Myr \citep[e.g.][]{2002scally}. By examining cluster dynamics of the massive and clustered star forming regions it is possible to set important constraints on models that would more effectively distinguish between these two theories. 

\citet[hereafter T09]{john} attempted to identify spectroscopic binaries that typically have significantly shorter separations through the multi-epoch spectroscopic monitoring of 1613 objects towards the ONC by searching for variability in the radial velocities (RV) that can be attributed to a presence of a companion. NGC 2264 has not yet been a subject to a systematic binary surveys, though \citet[hereafter T15]{ngc2264} did report on the multi-epoch spectra towards 695 objects towards this region.

T09 and T15 have previously analyzed the kinematic structure of the ONC and NGC 2264, building on the efforts of \citet{2006furesz} and \citet{2008furesz}. They found that in both of these regions there is a general agreement in the RV between stars and gas from which they have formed, which suggested that these regions are dynamically young, with ages of 1--2 crossing times. Surprisingly however, a significant fraction of the stars appeared to be blue-shifted relative to the gas, and there does not appear to be a significant number of red-shifted sources to balance the distribution. T15 showed that spectra of some of these blue-shifted objects found towards NGC 2264 exhibit Li I 6707\AA~ absorption. As this is an indicator of an extreme youth, while this is not a confirmation of the membership of the cluster, it suggests that sources that do contain Li I are at least casually related. No similar confirmation has been done for the sources in the ONC in T09.

In this paper, we revisit the published data supplemented by the more recent observations to identify a more complete sample of the multiple stars in the ONC and NGC 2264, as well as reexamine the kinematic structure of these regions. In Section \ref{sec:data} we present all the additional data taken since T09 and T15. In Section \ref{sec:analysis} we discuss the construction of the final catalog and the identification of binary stars. Section \ref{sec:multi} is focused on the specifics of the multiplicity in these regions, as well as fitting the orbits for select stars for which sufficient data were available, while Section \ref{sec:struc} looks at the stellar velocity distribution. Finally, in Section \ref{sec:conclusion} we summarize and discuss our findings.

\section{Data}\label{sec:data}

We reanalyzed all the spectra previously obtained by T09 for the ONC region and T15 for the NGC 2264 region
 (including several stars observed but not included in their published catalog) using Hectochelle \citep{hecto} and MIKE fibers\citep{mike1,mike2}. In addition to these data we include new observations from these instruments and now from the Michigan/Magellan Fiber System (M2FS)\citep{m2fs}.

\subsection{M2FS}

M2FS is a multi-object spectrograph on the Magellan Clay Telescope that is capable of both low resolution and high resolution echelle spectroscopy. Up to 256 targets can be observed over a 29' field of view. The fibers observing these targets are split into two independent yet identical spectrographs. M2FS fibers need to be plugged in manually into predrilled plates. The minimum allowed separations between fibers is 12''. A slitwidth of 180 $\mu$m yields a typical resolution $R \sim 20,000$.

We observed a total of 4 fields towards the ONC and 2 fields towards NGC 2264 in November 2013 and February 2014 (Table \ref{tab:inst}) using M2FS. We targeted a subset of stars from T09 and T15 sample that was previously thought to be varying and/or had multiple reliable detections. The Mg I filter was used covering the wavelength range of $\sim$5100---5210\AA.

In addition, we observed 3 fields towards the ONC in December 2014, also with M2FS but using H$\alpha$ and Li I filters, covering range of 6525---6750\AA. Since two separate orders are observed simultaneously, only 128 targets can be observed in this configuration. We obtained spectra mainly for the sources that have been originally identified by T09 as blueshifted relative to the gas in order to confirm their membership to the cluster via the presence of the Li I line, which can be used as an indicator of youth \citep{1997briceno}. Additionally, given available fibers, we observed objects that have been previously monitored and had $V$ fluxes between 12 and 13.5 mag.

Data were reduced using a custom Python code written by J. Bailey to merge the data and subtract the bias, and IRAF pipeline HYDRA to trace the orders, extract the spectra, calculate and apply the wavelength solution using a set of Th-Ar exposures, and perform the sky subtraction.

Spectra taken with Li I and H$\alpha$ filters were particularly affected by the strong nebular emission lines from S II (6717 and 6731\AA), N II (6549 and 6583\AA), as well as H$\alpha$. These features are always narrow, and appear in a conjunction with each other in any given spectrum. They would be present as the only features even in the spectra of stars that were too faint to be detected. All of these features were masked out in the final data product if they were observed on the visual examination. However, because the H$\alpha$ nebular line was often interfering with the one that should be observed due to stellar emission, often superimposed near the center of the line or barely offset from it, the masking process makes it impossible for us to reliably measure equivalent widths of H$\alpha$ for most spectra, and prevents us from detecting narrow stellar lines (both emission and absorption) in nearly all sources.

In addition to these narrow features, some spectra exhibited a very broad and strong emission-like feature at 6600 and 6725\AA~in H$\alpha$ and Li I orders, respectively, spanning $\sim$20\AA~in width. These features appeared at approximately the same pixel range in both orders and are thought to be caused by the Littrow ghosts from the optics. In the data taken with the Mg filter, a narrower and weaker feature appeared at 5181\AA~in the ``blue" spectrograph, and at 5187\AA~in the ``red" one. It is expected to have similar origins. In the cases where these features appeared as significant they were also masked out.

\subsection{Hectochelle}

In addition to T09 observations of the ONC made in 2007 with Hectochelle, data were acquired in 2008 and 2009 as well (Table \ref{tab:inst}). The multi-fiber echelle spectrograph Hectochelle on the MMT has a 1$^\circ$ field of view and can observe up to 240 targets simultaneously which can be positioned via robotic arms. The RV31 filter was used to cover wavelength range of $\sim$5150---5300\AA~with a typical resolution of $R \sim 35,000$. Data have been reduced using an IRAF pipeline developed by G. F\H{u}r\'{e}sz. A more detailed description of Hectochelle data reduction can be found in \cite{2006sicilia}.

\subsection{MIKE}

The stars that have been previously observed by T09 in ONC using MIKE fibers on the Magellan Clay telescope had an additional epoch observed in November 2008 (Table \ref{tab:inst}). MIKE consists of two independent spectrographs that can observe 128 fibers each. One of them was used to cover the wavelength of $\sim$5120---5190\AA, and the other one of $\sim$5140---5210\AA, at a resolution of $R \sim 18,000$. A description of the data reduction is available in T09.

\section{Radial velocity measurements}\label{sec:analysis}

All the data were processed through the IRAF package RVSAO \citep{rvsao} to extract the radial velocities (RV) from all the targets by performing cross-correlation against the synthetic stellar spectroscopic templates by \cite{munari}. As done by T09, all the templates had surface gravity $\log(g)=3.5$, effective temperature ($T_{eff}$) between 3500 and 7000 K in steps of 250 K, and solar metallicity. The previously reported spectra by T09 and T15 have also been re-correlated to achieve a homogeneous sample.

The default filtering parameters (low\_bin=5, top\_low=20, top\_nrun=125, nrun=255) were used during cross correlation to filter noise and large-scale structure in the spectra. However, rapidly rotating stars have broad and occasionally overlapping lines which would not be effectively processed with these parameters. For them different filtering parameters (low\_bin=3, top\_low=10) were used if the uncertainty from the revised correlation was no greater than 0.05 km s$^{-1}$ and the resulting measure of signal to noise $R$ value was greater,

\[ R=2^{-\frac{1}{2}}h\sigma_a^{-1}\]

\noindent where $h$ is the height of the peak of the correlation function, and $\sigma_a$ is the error estimated from the rms of the asymmetric component of the correlation \citep{1979tonry}.

In data taken in December 2014, the Li I and H$\alpha$ orders have been cross correlated separately. The H$\alpha$ line is the strongest feature in its order and the shape of the peak of the cross correlation is largely driven by the shape of this line. The H$\alpha$ line is usually much wider than the rotational velocity broadening, and because it is not photospheric, it can be affected by chromospheric motions. Additionally, because the center of the line was typically masked to remove scattered light, velocities obtained from this order are inherintly more uncertain than those obtained from the Li I order. If velocities obtained from both orders differed by less than the uncertainties added in quadrature, then average velocity and uncertainly ($v_{ave}$ and $\sigma_{ave}$) were calculated via the variance-weighted mean, and $R$ value was added in quadrature. This could only be done for 43\% of the sources. If no reasonable cross-correlation could be achieved from
the Li I order, H$\alpha$ velocities were used (9\% of the sources). 

Typical uncertainties for the individual RV measurements are 0.8 km s$^{-1}$ in the NGC 2264, and 1.2 km s$^{-1}$ in the ONC. Weighted average uncertainties for individual stars are 0.4 km s$^{-1}$ in the NGC 2264 and 0.7 km s$^{-1}$ in the ONC. It is possible that larger uncertainties in the measurements towards the ONC are due to the higher extinction.

When constructing the table of all the available measurements for all the sources, we retained only those measurements that had $R>3$ and $-100<$ RV $<100$ km s$^{-1}$. After that the time series of the measured velocities for each object was visually examined for inconsistent data. Common issues that were noted were as follows:

\begin{itemize}
\item Since the MIKE and M2FS fiber plates had to be drawn and plugged manually, occasionally a wrong star would be observed; this is found when the matched template for one or more is wildly different and would also exhibit velocity unlike the remaining observations of the same target.
\item Measurements with $3<R<6$ could be inconsistent with the remaining data for the target; this is likely a result of poor signal-to-noise that was not caught through an automatic filtering.
\item Many of the measurements taken on 12/02/2009 with Hectochelle appear to be contaminated by the moonlight; despite having a high $R$ value, these would typically have RV uncorrected for barycentric motion of $\sim0$ km s$^{-1}$.
\end{itemize}
Contamination from these measurements are among the main reasons for the discrepancy between the results presented in this paper and T09. These measurements have been removed from the final table and are not considered in any of the calculations. Any velocity measurement with $3<R<6$ that remained in the table was excluded from the following calculations described in this section as well, but they have been used only in visual examination of of the data to confirm the presence of the variability and in fitting the orbits of the identified binaries (Section \ref{sec:multi}). Beyond removing contaminating data, no velocity zero point offset was applied to the data taken on different days, as there appeares to be almost no systematic variability between different epochs. Median offset of individual measurements relative to the $v_{ave}$ of a given star within each epoch is typically within 0.3 km s$^{-1}$ and less than 1 km s$^{-1}$, which is consistent with the measured uncertainties.

We use the reduced $\chi^2$ as a measure of the consistency of the velocity in the time series. We identify systems as RV variable if they have reduced $\chi^2>16$ ($\sim 4\sigma$). In all, there are 2057 sources with at least a single velocity measurement with $R>6$, of which 1154 are found towards ONC and 903 towards NGC 2264 (Table \ref{tab:single}). A total of 130 sources have been identified as RV variable, with 79 toward the ONC region and 51 toward NGC 2264 (Table \ref{tab:binary}). Individual measurements of all the non-variable sources it is reported in Table \ref{tab:single}. All the sources identified as RV variable are listed in the Table \ref{tab:binary}.

\section{Multiplicity}\label{sec:multi}
\subsection{Measured multiplicity fraction}

Out of 137 sources originally identified by T09 as RV variables in the ONC region, we can confirm only 15 as such in our final catalog. The remaining sources exhibited either little to no change in velocity or had variable velocity measurements that were of low significance.
Some of those sources could still be multiple systems, but we do not consider them further here due to our
stricter limits for significance to avoid false positives. Upon closer examination, those sources that have been previously identified as double-lined binaries by the presence of the second peak in the correlation either could not be confirmed as such, or they were flagged as variable by our method. For this reason we focus only on the 130 sources that we can only identify as RV variables with either new or reanalyzed spectra.

We consider two possible causes of the RV variability. Firstly, it could be due to the orbital motion of the muliple systems. Secondly, changing radial velocity could be a result of RV jitter due to spots on the surface of magnetically active stars. The typical effect of jitter in the main sequence stars is on the order of a few m s$^{-1}$ \citep{2015Hillenbrand}, but it could be on the order of $\sim$ 1 km s$^{-1}$ in the pre-main sequence stars \citep{2013donati, 2014donati, 2015donati}.

Given the fact that typically RV variable sources had significant fluctuation from the $v_{ave}$, RV jitter can account only for a handful of sources. We identify sources as multiple if they have at least one measurement $|v_{ave}-v|>4$ km s$^{-1}$. A total of 113 sources satisfy this. Remaining 17 sources fall below this threshold; while they do exhibit RV variability, we cannot confirm that it is due to an orbital motion within a multiple system. Because of this we exlcude them from any calculations involving multiplicity fraction.

\begin{figure*}
 \centering 
			\subfigure{\includegraphics[width=0.45\textwidth]{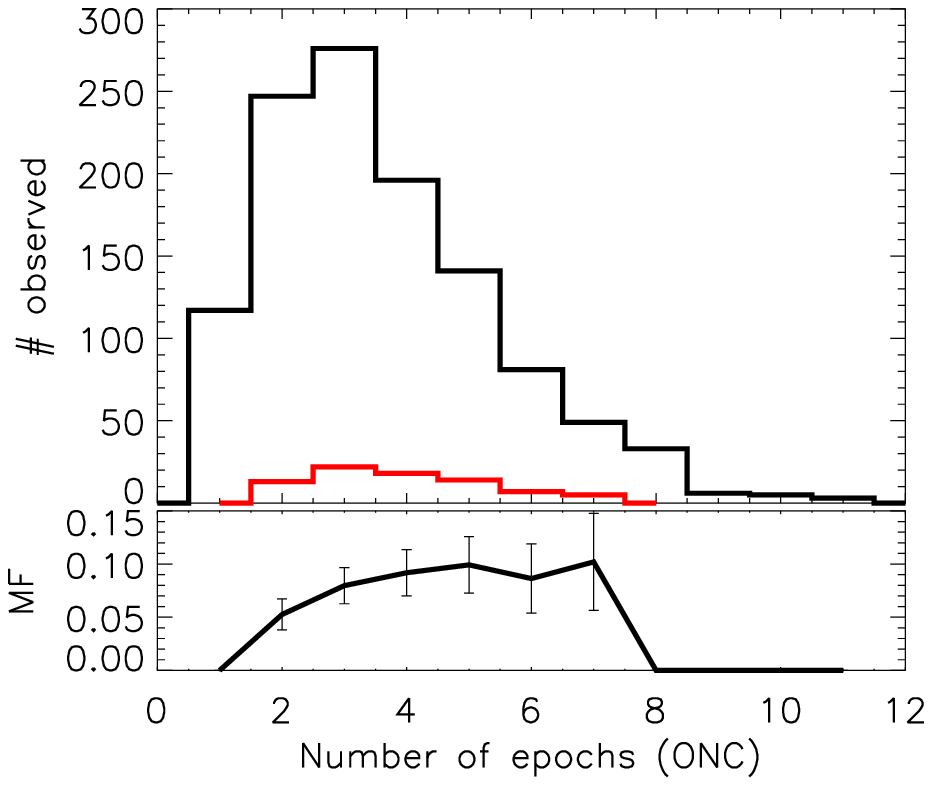}}
 			\subfigure{\includegraphics[width=0.45\textwidth]{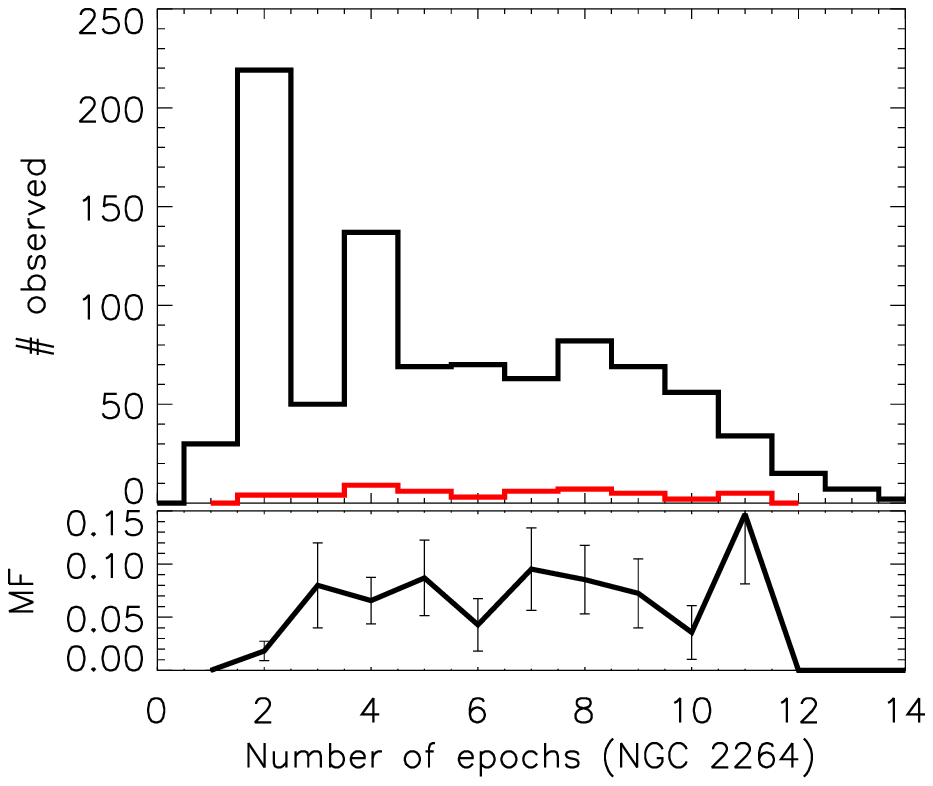}} 
 \caption{Number of observations made for all the objects in the sample, shown in black. Red shows only sources with $\chi^2>16$. The bottom panel shows the ratio of the number of sources with $\chi^2>16$ per number of all sources with a given number of epochs observed. \textbf{Left:} ONC. \textbf{Right:} NGC 2264.}\label{fig:nobs}
\end{figure*}

The velocity curves for most of our sources are very undersampled, and so we could fail to detect real RV variables. We consider the minimum number of epochs required for a guaranteed detection of an RV varying system to be 3. With only 2 measurements it is possible to miss a binary system due to the coincidental timing between the observations. However with 3 or more epochs of data the fraction of the number of systems identified as multiples to the total number of stars remains relatively unchanged (Figure \ref{fig:nobs}).

We attempted to identify false positive sources that were flagged as multiples if they had only single discrepent measurement from the mean. To do that we calculated the reduced $\chi^2$ thowing out the single most variable RV measure from $v_{ave}$ and identified sources where this revised reduced $\chi^2<16$. We required that (a) all the sources flagged by this method have at least 4 measurements with $R>6$, since otherwise lack of the detected variability could be due to poor sampling; (b) at least 4 epochs during which the measurements were taked are separated in time by more than 2 days, as such short separations in time cannot detect variability due to orbital motion with orbits longer than a few days; and (c) only a single measurement has a discrepent RV, including $R<6$ measurements if they have comparable variability from $v_{ave}$. This identified six sources -- RV 10, 138, 929, 1372, 1496, and 1660. In cases of RV 1372, 1496, and 1660, we provide a detailed orbital fits (Section \ref{subsec:fit}). These fits provide little doubt in the nature of these sources as muliple systems; however, without a single strongly deviating RV they would not have been identified as such based on the data in this paper. For this reason, while the remaining three sources (RV 10, 138, and 929) are flagged in the Table \ref{tab:binary} as possible false positives, we include them in the flollowing calculations.

The observed multiplicity fraction (defined as the overall number of the multiple systems, hereafter MF) within the RV dataset is $8.0\pm 1.2$\% towards the ONC, and $6.7\pm 1.1$\% towards NGC 2264 if we include all the sources in the Tables \ref{tab:single} and \ref{tab:binary}. However, if we require $v_{ave}$ converted from heliocentric to local standard of rest (lsr) reference frame \citep{1986kerr} to range from -5 to 20 km s$^{-1}$ for both single stars and binaries to limit the contamination from the sources that are not the members of these clusters (T15, also more in Section \ref{sec:struc}), MF becomes $5.8\pm 1.1$\% for ONC (30 multiples out of 518 stars observed), and $5.3\pm 1.2$\% for NGC 2264 (21 out of 397). Uncertainties were obtained as $N_{multiple}^\frac{1}{2} N_{total}^{-1}$.

\subsection{Comparison to the field}

To compare the MF that we observe towards ONC and NGC 2264 to that observed in the field, we ran a Monte-Carlo simulation producing a synthetic field population that would be consistent with distribution of binary properties of the nearby G dwarfs \citep{2010raghavan}. For each of the stars in our sample that had $-5<v_{lsr}<20$ km s$^{-1}$ and $>$3 detections (i.e., those from which the MF was measured) we generated randomly configured systems (including both single and multiple) that had the same dates of the observation and same uncertainties as the data. We then ran the same detection test that would identify binary systems from the generated population producing a MF for a single test case. This process was repeated 1000 times. Average MF from all test cases both in ONC and NGC 2264 regions was determined separately. Uncertainty was determined from 1$\sigma$ dispersion in the generated MF between 1000 test cases.

For the ONC there are 518 systems that make up the population from which the MF was estimated. Of these, $\sim$120 stars have uncertainties too large to be detected as variable in any of the runs. The average MF of the synthetic population is $4.8\pm0.9$\%. NGC 2264 had a total of 397 systems of which $\sim$70 cannot be identified as multiples in any run, and the average MF of $6.1\pm1.2$\%.

\begin{figure}
\centering
\subfigure{\includegraphics[width=0.2\textwidth]{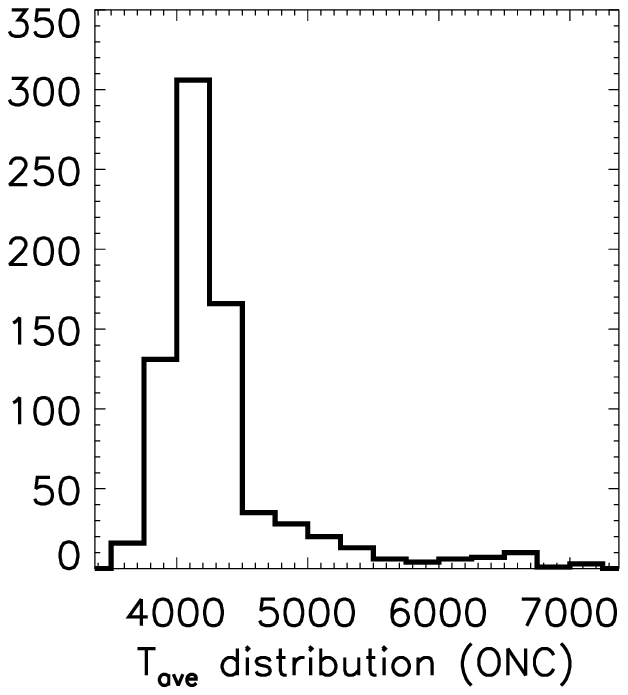}}
\subfigure{\includegraphics[width=0.2\textwidth]{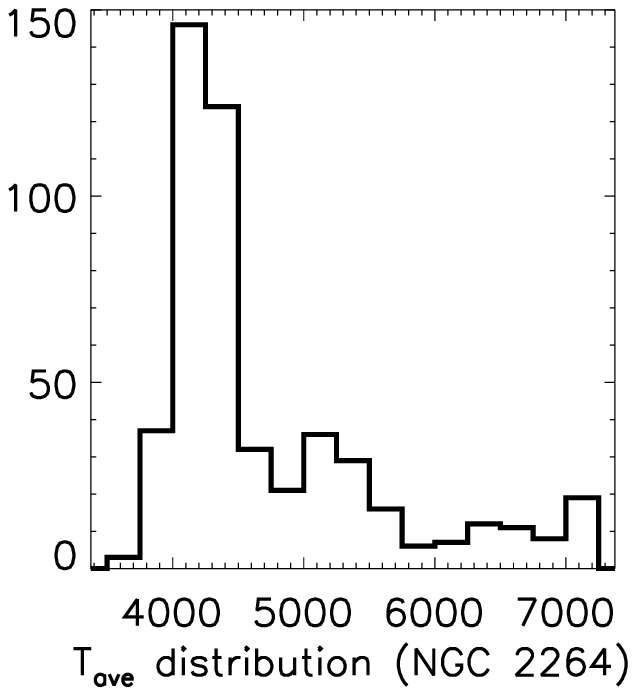}}
\caption{Average effective temperature disribution observed towards stars in ONC and NGC 2264.} \label{fig:temp}
\end{figure}

The estimation of the expected MF makes an implicit assumption on the mass of the primary stars. To simulate the population of the field stars to be consistent with what was measured by \cite{2010raghavan}, 1 M$_\odot$ primaries were chosen. However, the masses of stars monitored in these clusters expected to be significantly lower. The typical effective temperature template matched towards these stars is 4000 K (Figure \ref{fig:temp}). From this we can estimate them to be K stars, with their typical mass  on order of 0.7 M$_\odot$ \citep{2015baraffe}. Stellar multiplicity varies strongly with stellar mass, and lower mass stars have been found to have lower MF than higher mass stars \citep{1992fischer, 2013duchene}. G stars have typically been used for comparison because surveys of their orbital parameters have been presently by far the most comprehensive.

In addition, because the mass of the primary is lowered, the distribution of the masses of the secondary is also cut off at a smaller values. Because of this, the difference in peaks of the RV fluctuations due to the presence of a companion would become smaller as well; fewer systems would be detected. By generating a population with identical orbital parameters around 0.7 M$_\odot$ instead we infer a MF of $4.5\pm0.9$\% for ONC and $5.8\pm1.2$\% for NGC 2264.

The dispersion in the MF generated with identical orbital parameters between separate runs (all consisting of 1000 test cases) observed with the cadence and uncertainties set by the same population is typically $<0.1$\%. To estimate the systematic effects of due to a potentially different distribution of periods and overall number of binary systems in these clusters as opposed to what was previously found in the field, we varied one parameter at a time and looked for a difference in MF. Varying the underlying binarity fraction by 2\% \citep[1$\sigma$ values quoted by][]{2010raghavan} typically changes the extracted MF by 0.2\%. No uncertainties on the orbital parameters were made available, although varying the peak of the period distribution by 0.1 $\log P$ (where $P$ is measured in days) produces an MF that is different by 0.5\% when the average period is decreased from 293 to 233 years, and by 0.3\% when the period is increased to 369 years. Varying the standard deviation of the period distribution by 0.1 $\sigma_{\log P}$ changed MF by 0.4\%. Changing the mass ratio and eccentricity from uniformly distributed to those that are described by \citet{1991duquennoy} decreases MF by 0.5\% and 0.4\% respectively.

\begin{figure}
 \centering 
			\subfigure{\includegraphics[width=0.2\textwidth]{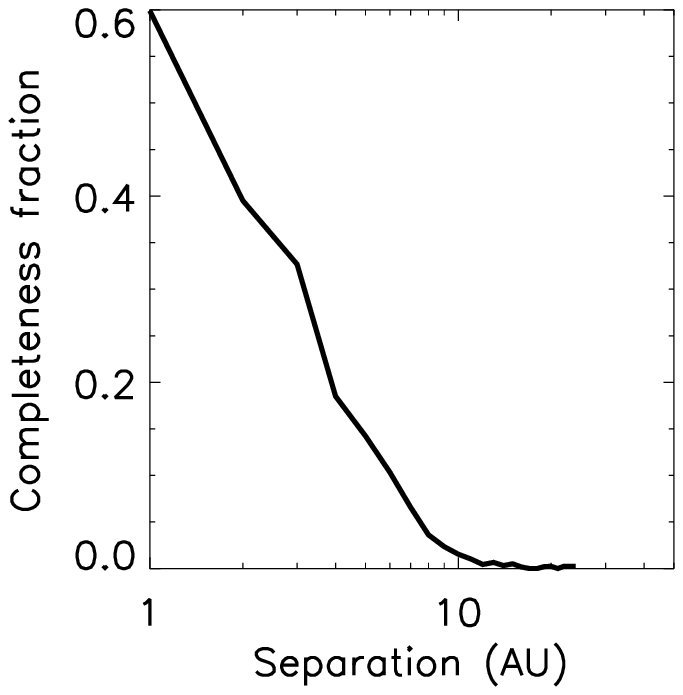}}
 			\subfigure{\includegraphics[width=0.21\textwidth]{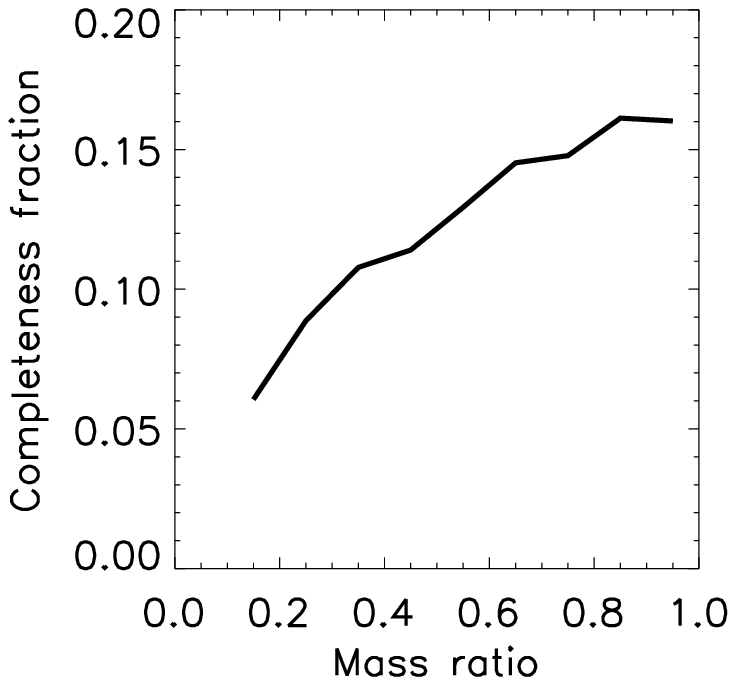}} 
 			\subfigure{\includegraphics[width=0.2\textwidth]{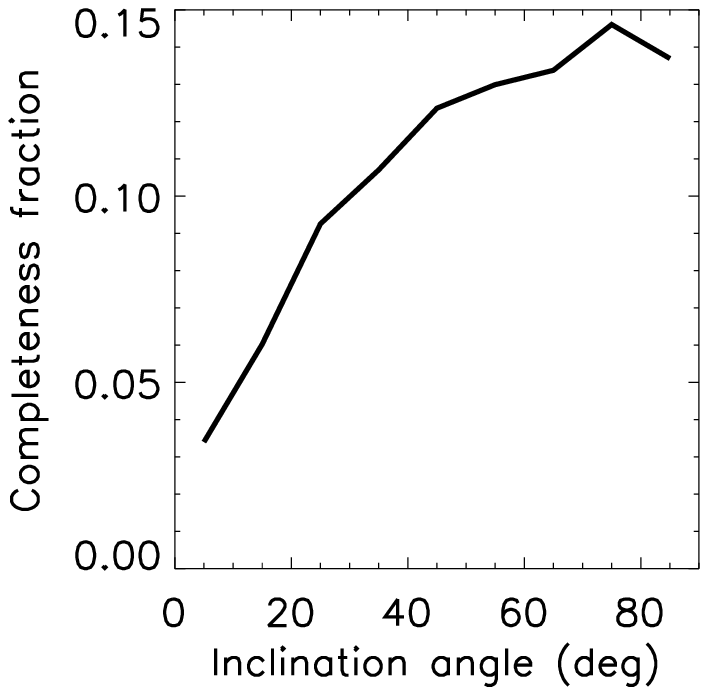}} 
 \caption{Completeness limits of the detection of the multiple systems depending on their orbital parameters. }\label{fig:complete}
\end{figure}

To determine the completeness limits which we probe with these observations, we recorded orbital parameters from all the generated binary systems and determined a fraction that would be detectable relative to the all binaries that satisfy the specific orbital parameters (Figure \ref{fig:complete}). Unsurprisingly, most easily detectable systems have short orbital period, comparable masses between the primary and the secondary, and an edge on orientation. It is possible to detect only 60\% of all the systems with separations $<1$ AU because either the other orbital parameters make a detection difficult, or the uncertainty in the measurements that were applied to the generated velocity curves were too large for a reliable detection. Beyond the separations of 15 AU there are almost no systems that could be detected as a binary based on their RV variability. Combining all the possible separations and inclination angles, only 15\% of multiple systems can be detected for stars with the mass ratio on the order of unity (in this case both stars have mass of 1 M$_\odot$), however, this rapidly decreases to $\sim$5\% for the companions with only 0.2 M$_\odot$.

\subsection{Orbital parameters}\label{subsec:fit}

Individual cross-correlations of all systems identified as binaries were visually examined to determine whether or not it is possible to see a second peak due to the presence of the second star. We required that any star flagged as a double-lined binary exhibited multiple peaks or a skewed correlation function in at least two epochs to minimize spurious detections. There were a total of 15 of such systems, of which 10 were found towards NGC 2264 and 5 towards ONC. In epochs where it was possible, a Gaussian was fitted to both peaks to find the velocities of both components. We assigned uncertainties to these measurements of 2 km s$^{-1}$, limited by the resolution of the extracted cross-correlations function. All of these measurements are included in Table \ref{tab:binary}.

Out of 130 sources flagged as binaries in this paper, six single-lined binaries have $\geq 10$ RV measurements, and two double-lined binaries have $\geq 9$ measurements. We attempted to fit the velocity curves of these sources as they contained a sufficient number of measurements and redundancy on the measurements to obtain a unique solution. Because sources in NGC 2264 have been monitored more frequently than ONC (Figure \ref{fig:nobs}), all of these 8 sources are found towards NGC 2264.

To fit the orbits, the IDL package RVFIT \citep{rvfit} was used, which fits the following quantities -- $P$ (the orbital period), $T_P$ (the time of periastron passage), $e$ (the eccentricity), $\omega$ (the argument of the periastron), $\gamma$ (the systemic velocity of the system), and $K_1$ (the amplitude of the radial velocity fluctuation) and $K_2$ (the amplitude of the radial velocity fluctuation for the second star, if the system in question is a double-lined  spectroscopic binary.) The derived quantities are the semimajor axis $a_1\sin i$ and the binary mass function $f$, and it can distinguish between contributions of individual stars for double-lined systems.

RV 1768 showed the second peak in the correlation in 7 out of 11 epochs, and both peaks exhibited a similar strength. The remaining 4 epochs had measurements very close to the mean of the remaining measurements. Because we are unable to distinguish contributions from the individual components, there could be a larger spread in the velocity due to line blending; we assign uncertainty to those measurements of 5 km s$^{-1}$.

\begin{figure}
\epsscale{1}
\plotone{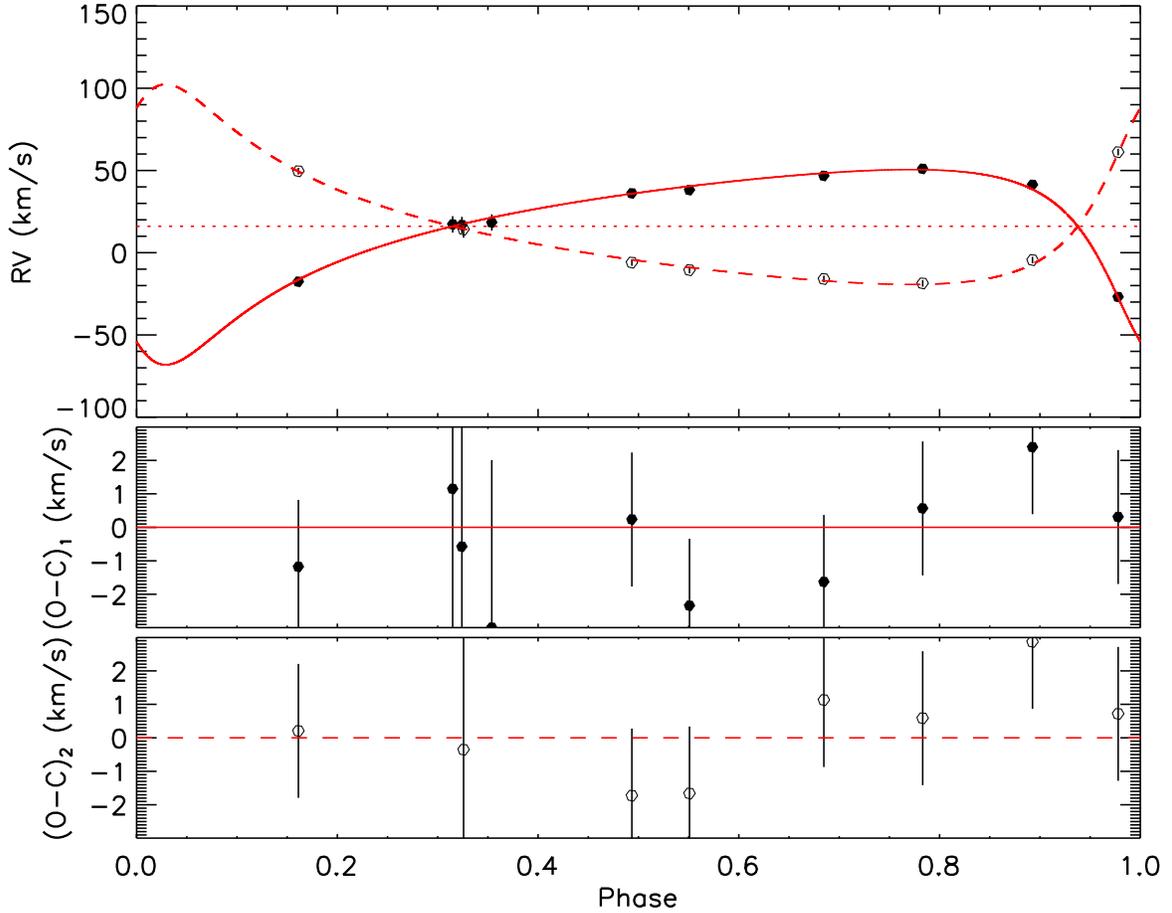} 
\caption{Orbital fit for RV 1768\label{fig:rv1}}
\end{figure}

As RV 1768 system appeared to have almost equal mass in both components, there was some difficulty in distinguishing which of the two lines in a given epoch belonged to which stars in case of the resolved measurements, and the velocity of which star was the most dominant in case of the unresolved measurements. All the perturbations of line combinations were attempted to be fitted and examined by eye - the combination which provided the best fit is listed in Table \ref{tab:binary} and the resulting fit is shown in Figure \ref{fig:rv1} and Table \ref{tab:dbpeak}. The masses of the individual components (M~$\sin^3 i$) are $\sim 0.25$ M$_\odot$. Given that the spectra for this system are best fitted by a $\sim$4000 K template, this is a reasonable fit requiring only modest inclination angle.

\begin{figure}
\epsscale{1}
\plotone{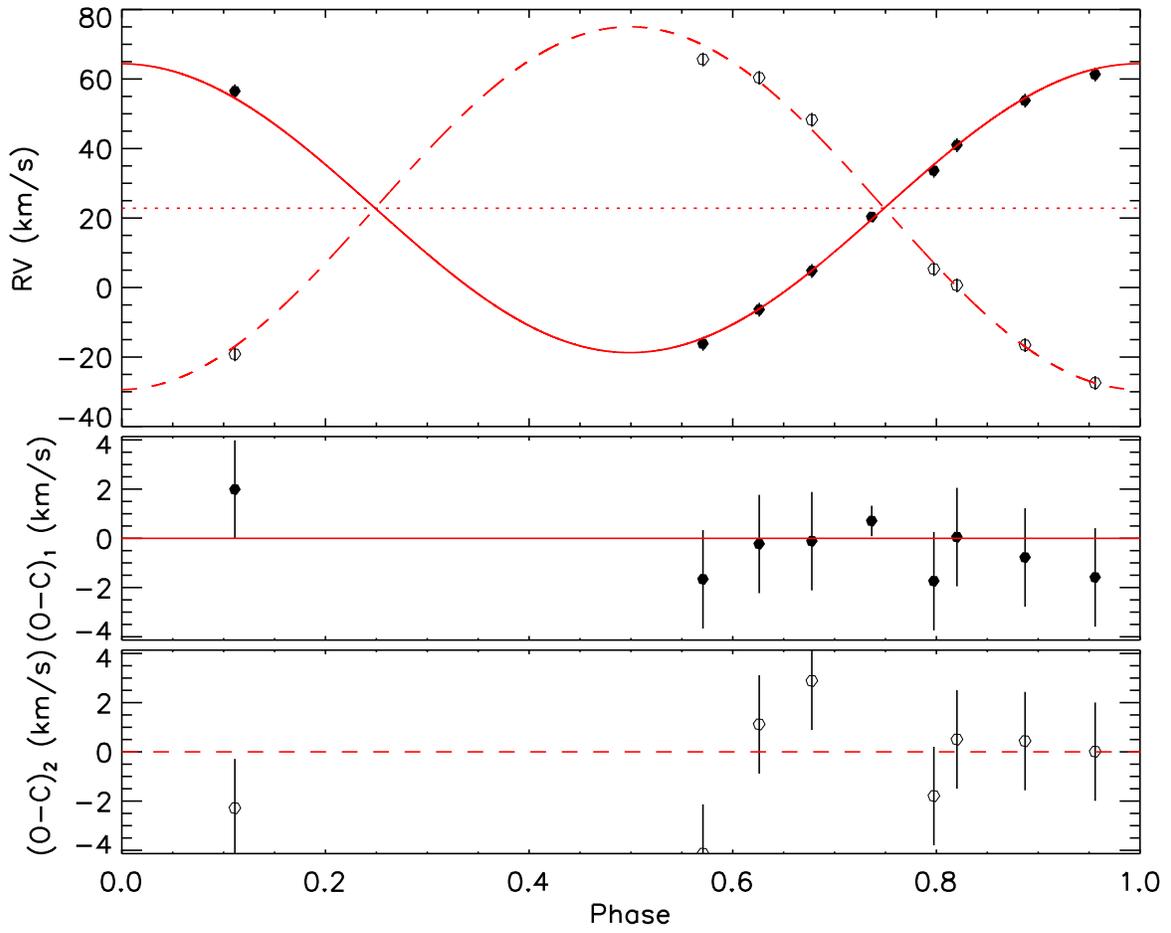} 
\caption{Orbital fit for RV 1659\label{fig:rv8}}
\end{figure}

RV 1659 had a double-peaked correlation function in 8 out of 9 epochs, with one component clearly dominating over the other. It was best-fitted by a circular orbits, period of 15.3 days, and M~$\sin^3 i$ of 0.73 and 0.58  M$_\odot$ (Figure \ref{fig:rv8}, Table \ref{tab:dbpeak}). For orbits with no eccentricity, argument of the periastron carries no meaningful information.

\begin{figure}
\epsscale{1}
\plotone{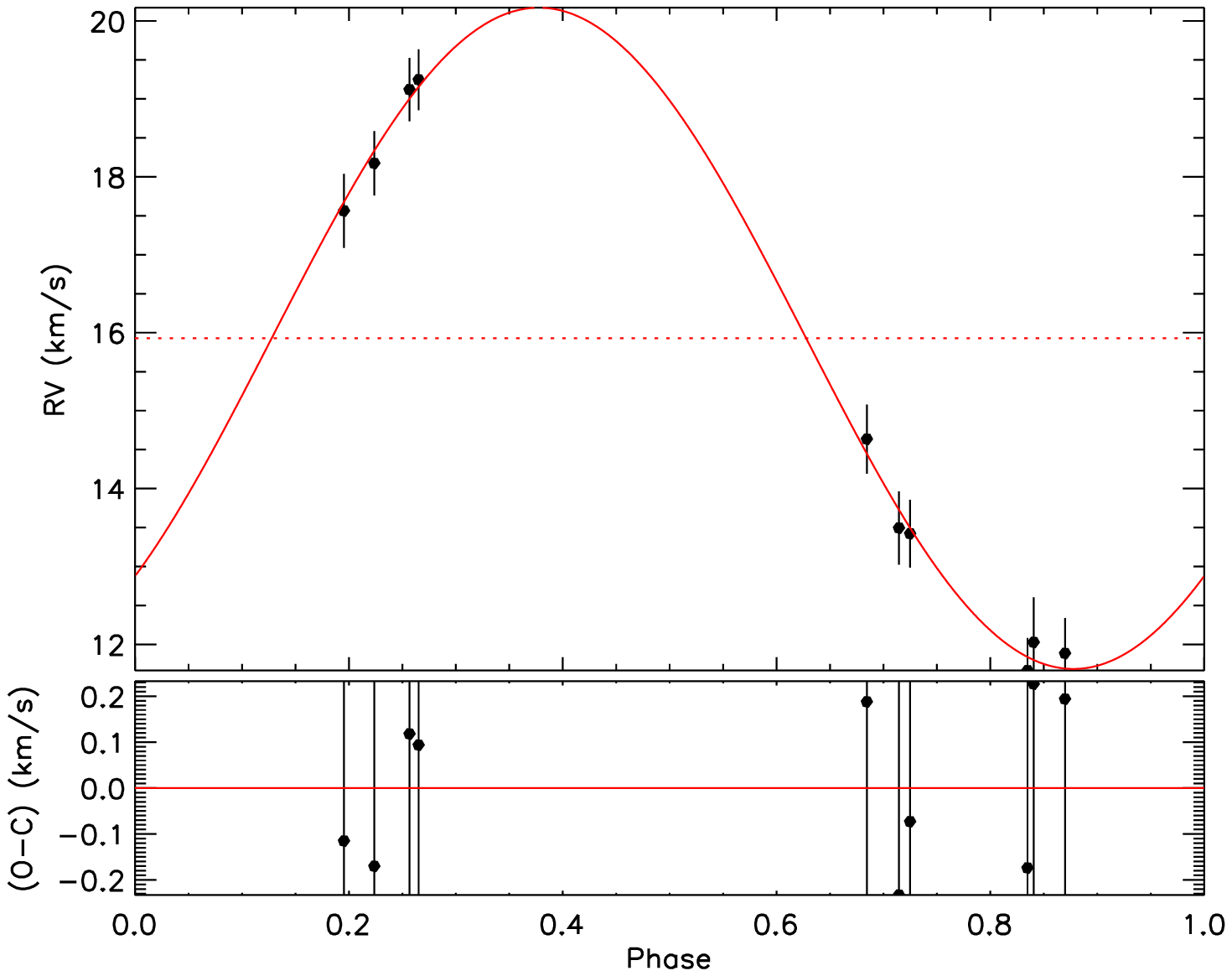} 
\caption{Orbital fit for RV 1166\label{fig:rv6}}
\end{figure}

\begin{figure}
\epsscale{1}
\plotone{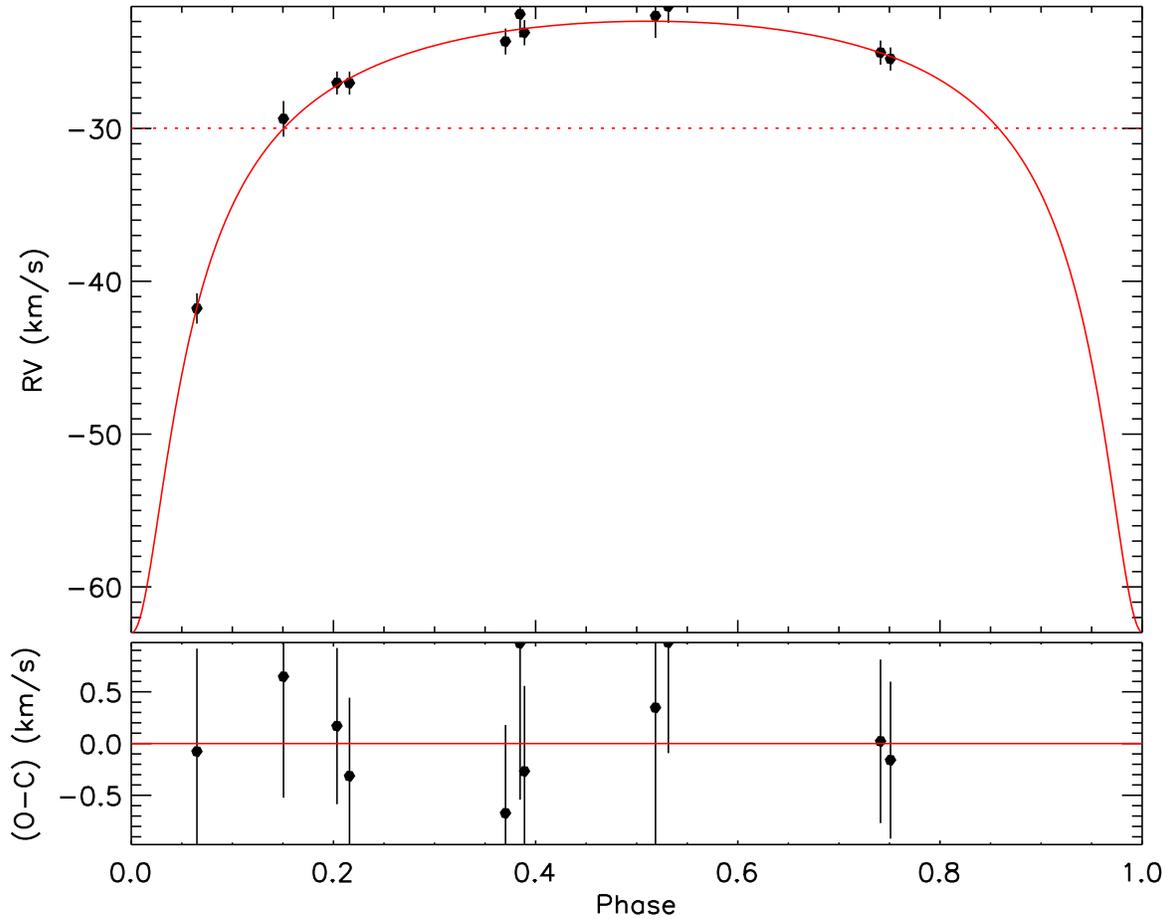} 
\caption{Orbital fit for RV 1372\label{fig:rv2}}
\end{figure}

\begin{figure}
\epsscale{1}
\plotone{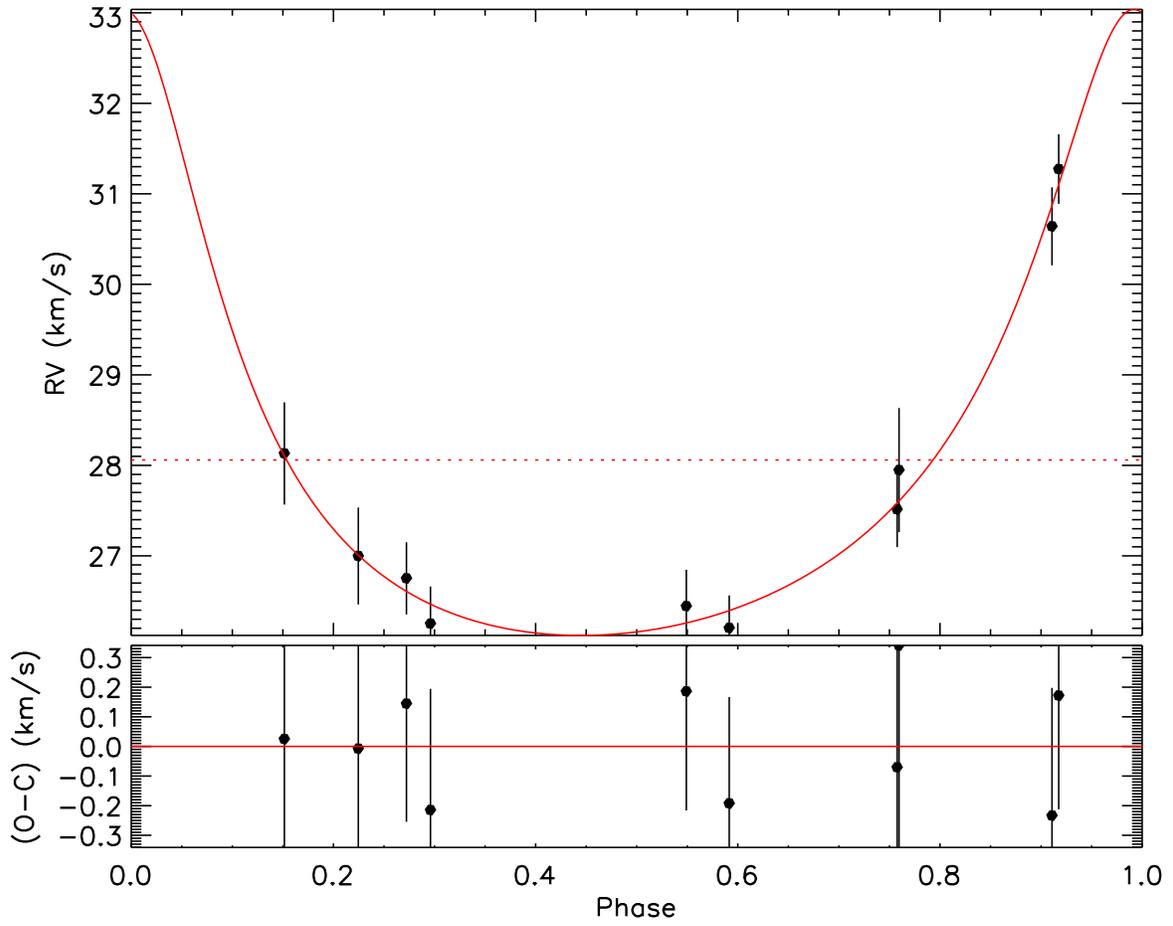} 
\caption{Orbital fit for RV 1496\label{fig:rv5}}
\end{figure}

\begin{figure}
\epsscale{1}
\plotone{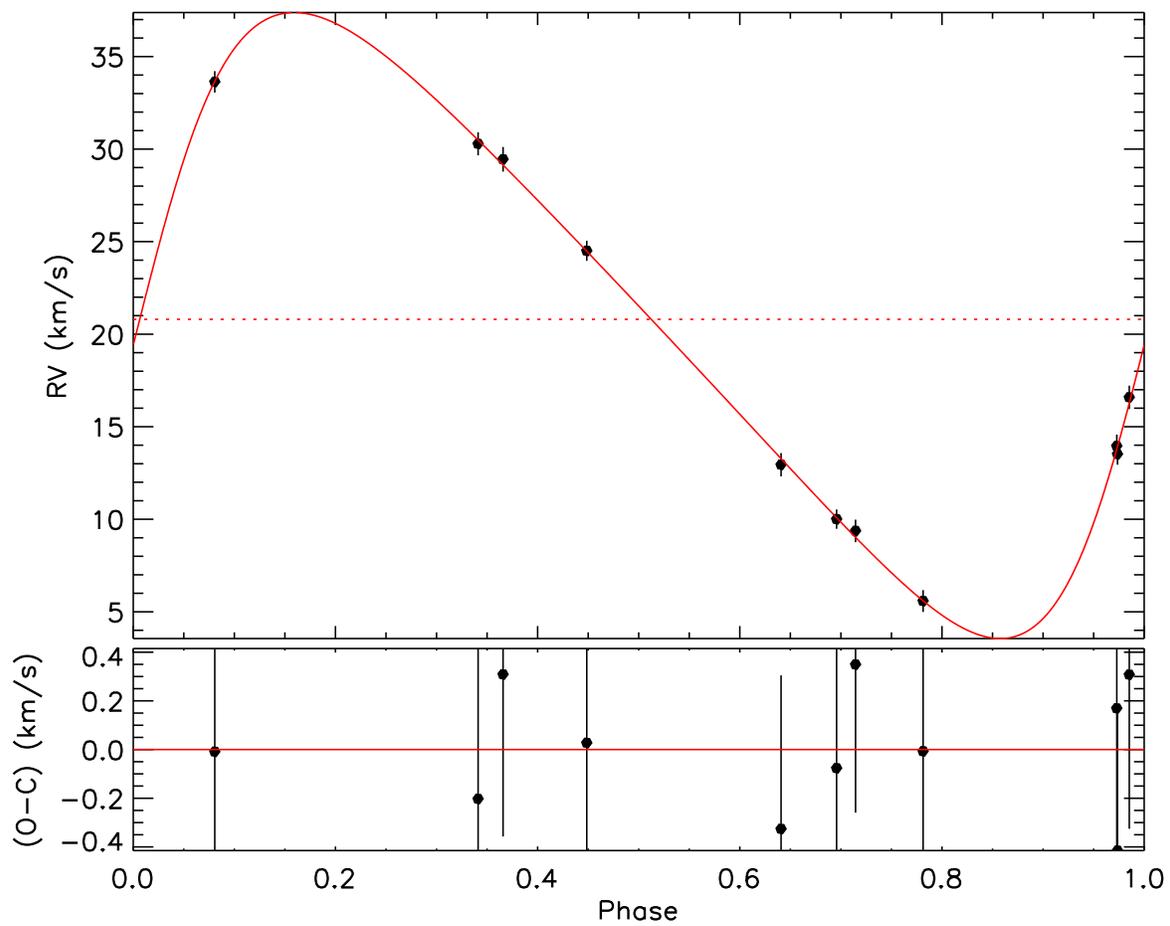} 
\caption{Orbital fit for RV 1550\label{fig:rv4}}
\end{figure}

\begin{figure}
\epsscale{1}
\plotone{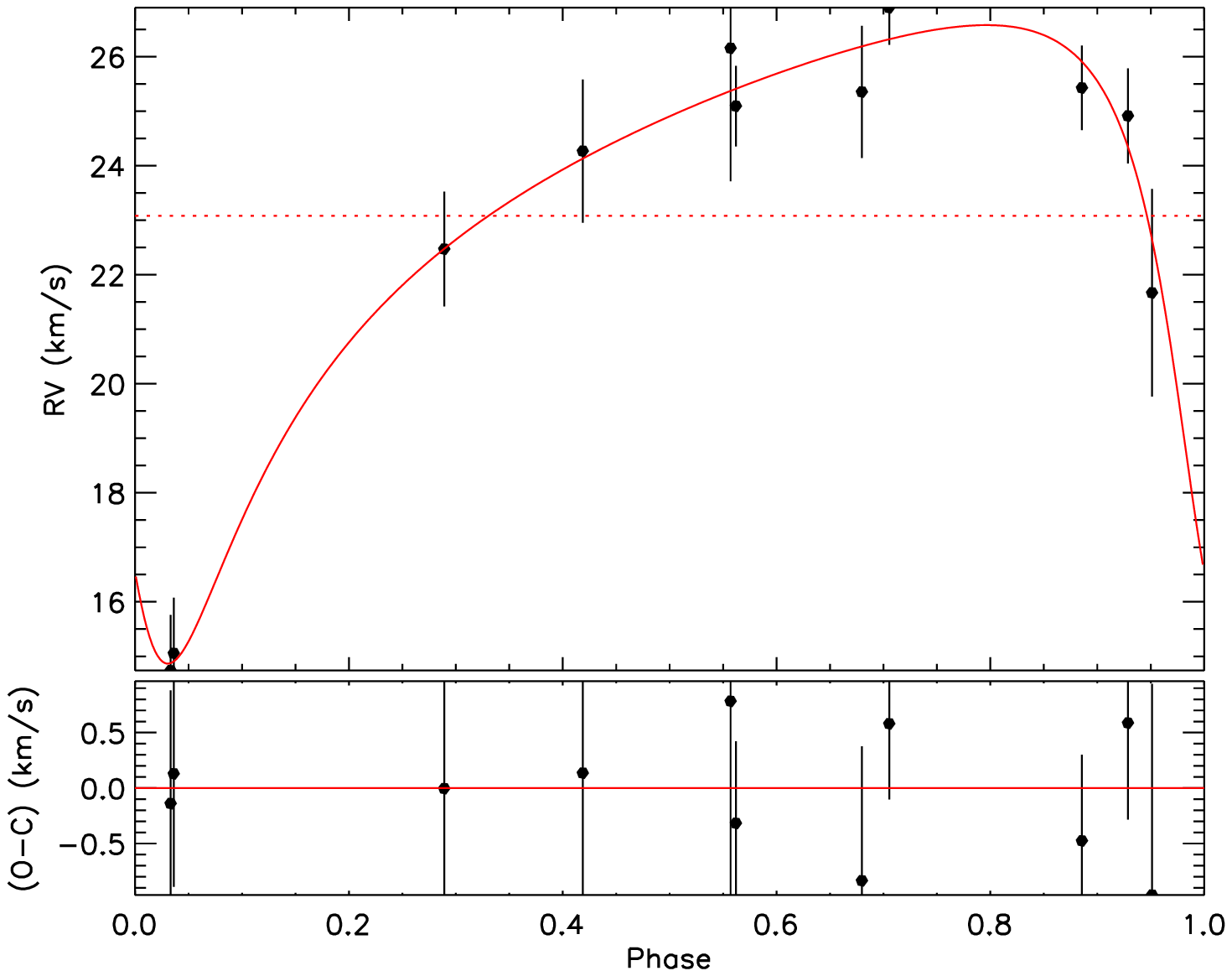} 
\caption{Orbital fit for RV 1660\label{fig:rv3}}
\end{figure}

\begin{figure}
\epsscale{1}
\plotone{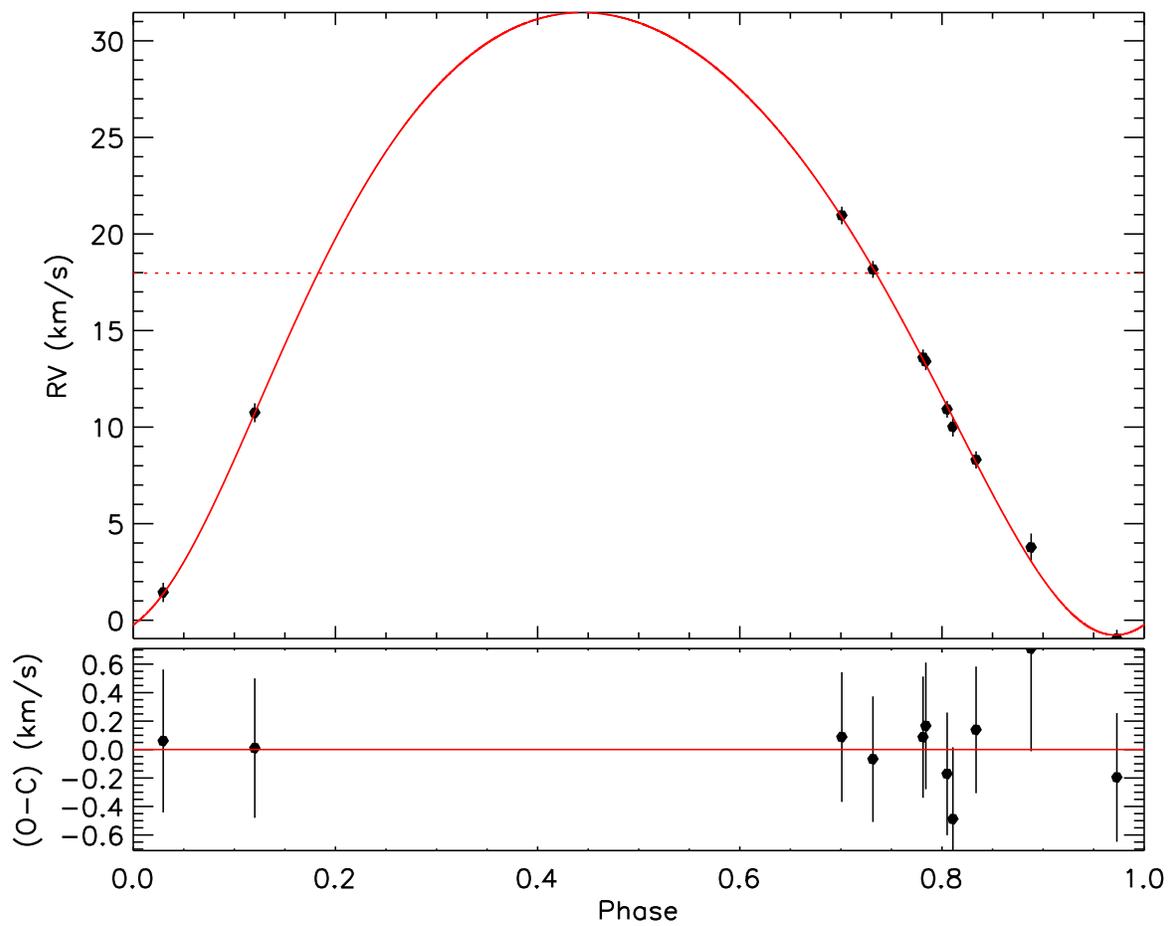} 
\caption{Orbital fit for RV 1753\label{fig:rv0}}
\end{figure}

Fitting information on the six single-lined systems is presented in the Table \ref{tab:speak} (Figures \ref{fig:rv6}--\ref{fig:rv0}). Based on the characteristic heliocentric velocity of RV 1372 of -32  km s$^{-1}$ (v$_{lsr}\sim-47$ km s$^{-1}$), it is unclear whether or not it is a member of NGC 2264. However, this system is typically best-fitted by a 4250 K template and has near infrared fluxes of  J=12.270, H=11.718, K=11.464 \citep{2mass}. While there is some contamination in the fluxes from the companion, they are in a good agreement with those of other binary systems that follow the same templates (e.g. RV 1550, a system for which the orbital fit is also available and that has systematic velocity that is consistent with the cluster mean, has  J=12.262, H=11.566, K=11.423), making it likely that RV 1372 is not a foreground or background system, but rather that it was ejected from the cluster.

RV 1753 have been previously monitored for change in radial velocities by \citet{2013karnath} over the period of 20 years with 35 epochs. All the fitted orbital parameters from that study are in excellent agreement to the fits presented in this paper.

\section{Velocity structure}\label{sec:struc}
\subsection{ONC}\label{subsec:onc}

\begin{figure}
\epsscale{0.9}
\plotone{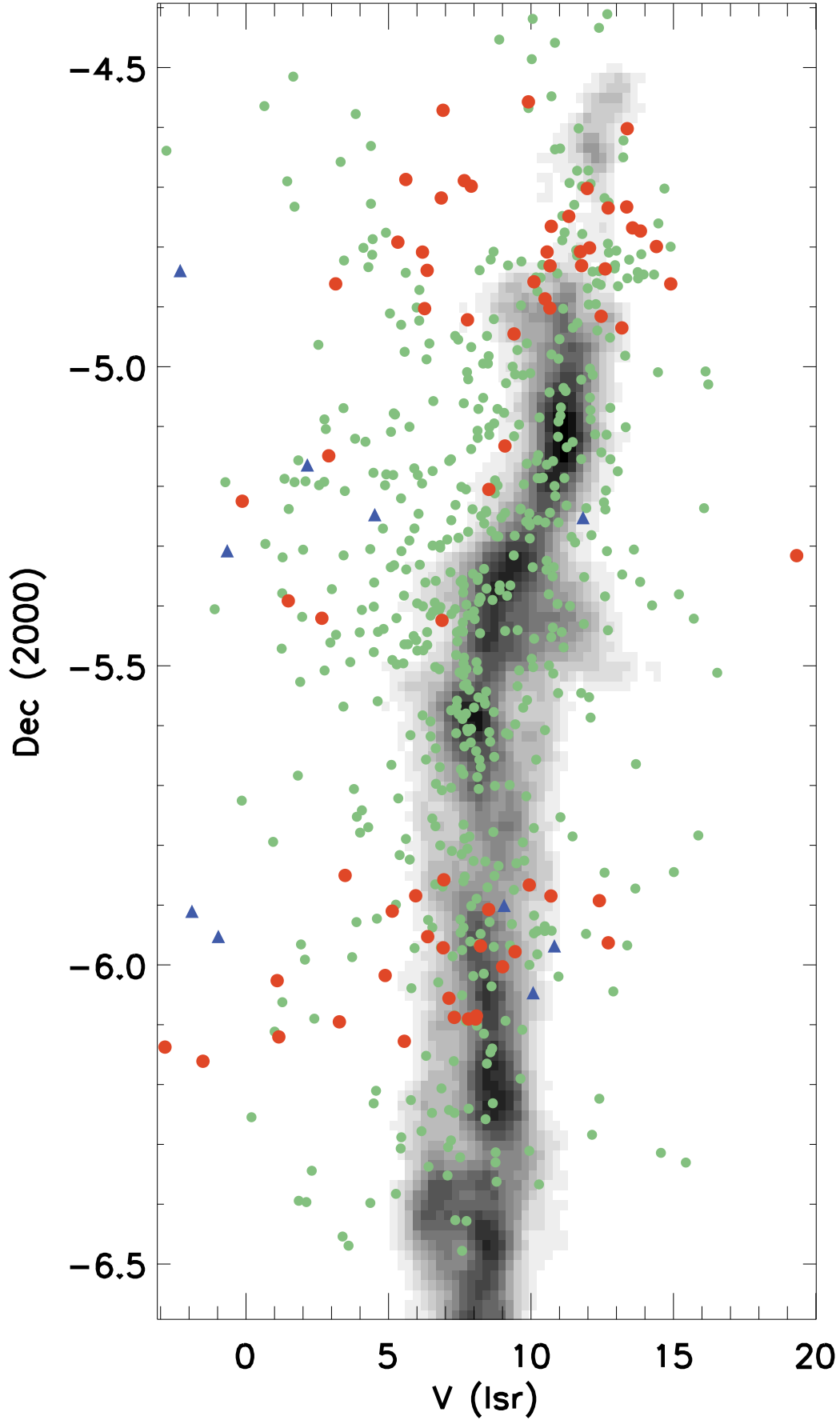} 
\caption{Position-velocity diagram for the ONC region, summed in right ascention. $^{13}$CO map from \citet{1987bally} is plotted in the background in grayscale. All the overplotted data points are non-binary sources that were observed in at least 3 epochs. Orange circles show sources where Li I has been detected, blue triangles show those that have been surveyed for the presence of Li I, but it was not detected. Green dots are all the remaining sources for which no Li I information is available.}\label{fig:co}
\end{figure}

\begin{figure}
\centering
\subfigure{\includegraphics[width=0.2\textwidth]{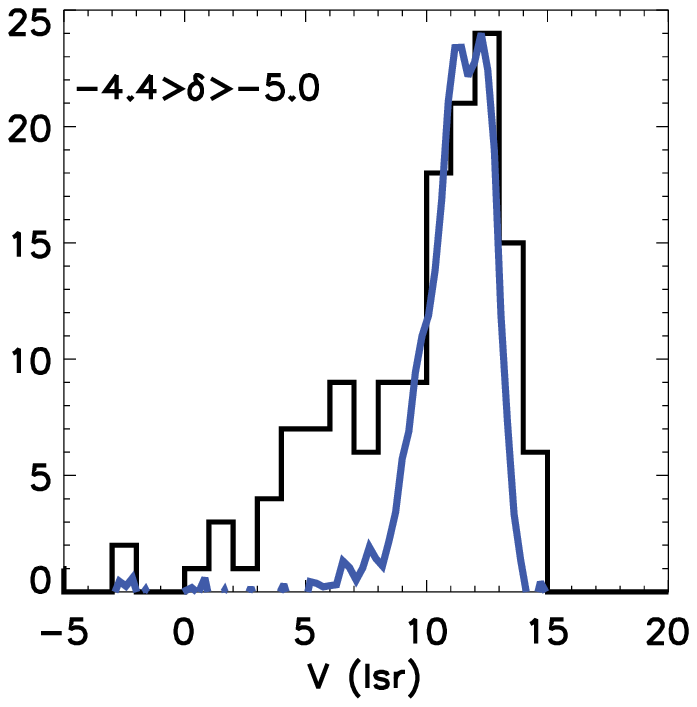}}
\subfigure{\includegraphics[width=0.2\textwidth]{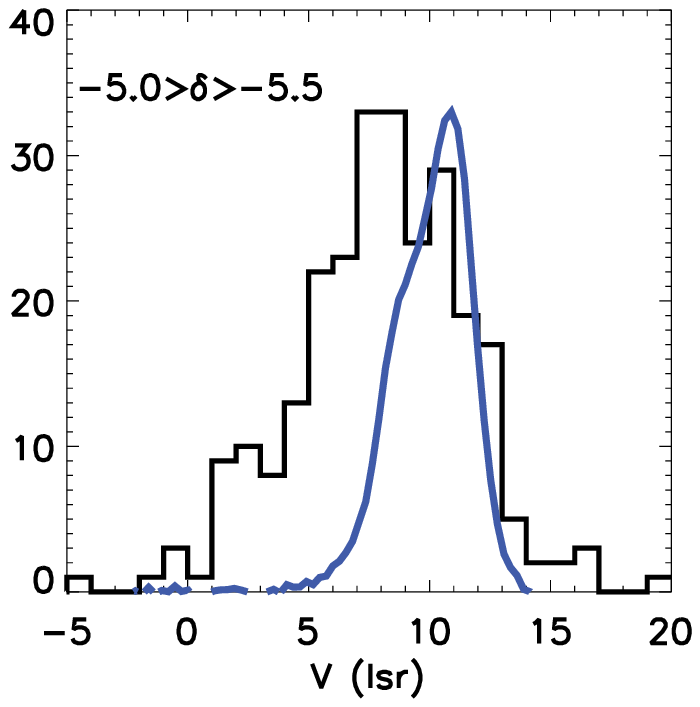}}
\subfigure{\includegraphics[width=0.2\textwidth]{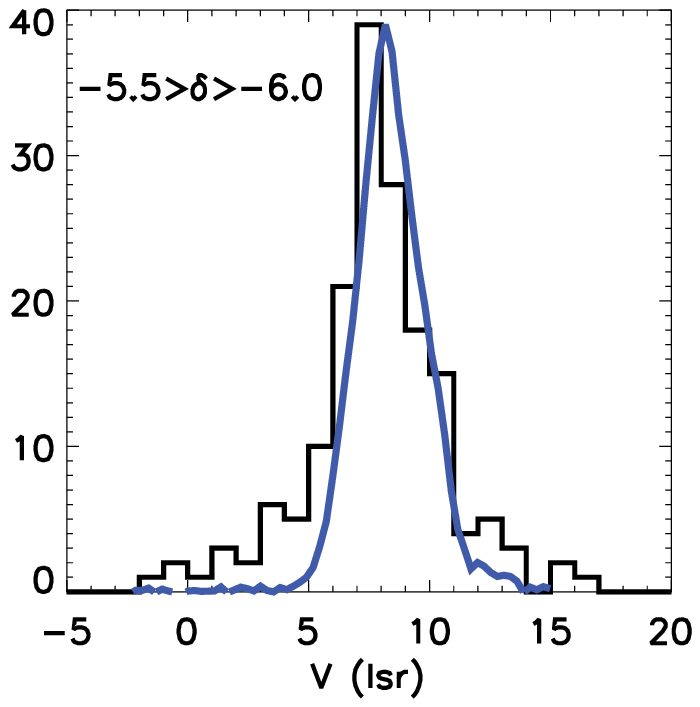}}
\subfigure{\includegraphics[width=0.2\textwidth]{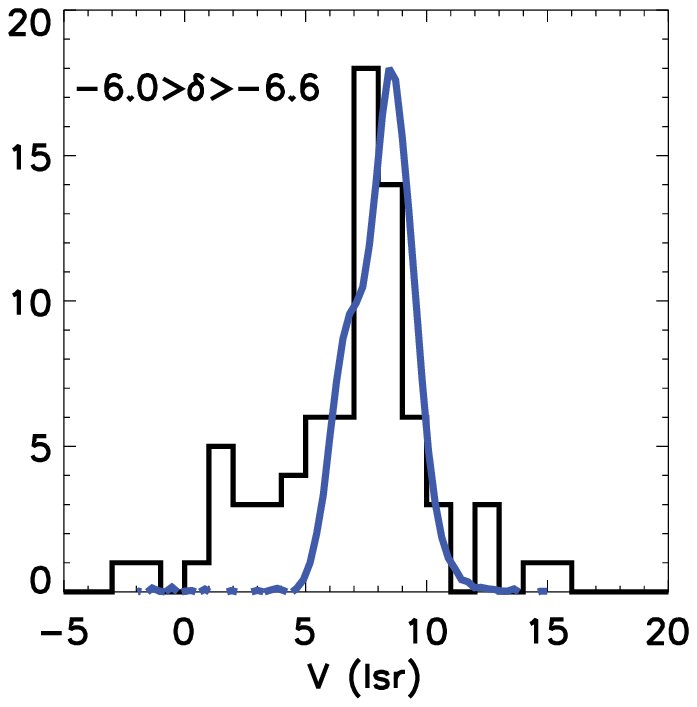}}
\caption{In black - distribution of of velocities of stars plotted in Figure \ref{fig:co} at four declination cuts. In blue - summed distribution of the $^{13}$CO at those declinations, scaled to the peak of the histogram.} \label{fig:cohist}
\end{figure}

Some of the analysis of the ONC region performed in T09 was affected by the contamination from moonlight, and the lowest signal-to-noise data also added scatter. While the overall conclusions of T09 remain unchanged, the sample presented in this paper has the velocity measurements of the higher quality. For this reason once again we look at the relation between stellar radial velocities and $^{13}$CO gas (Figure \ref{fig:co}, \ref{fig:cohist}). All the velocities in the plots are in the kinematic LSR reference frame to match that of the gas \citep{1987bally}. To confirm that no binary stars contaminate the sample, we required no variability in velocities with at least 3 epochs of observations. Unlike in T09, the peak of the RV distribution for stars follows the gas with no offset and a comparable velocity dispersion of $\sim$2.5 km s$^{-1}$. The only exception to this is the $-5.5^\circ<\delta<-5.0^\circ$ range, which coincides with the location of the inner ONC regions such as Trapezium and OMC 2/3. A recent paper by \citet{2015dario} that measured RV from the infrared spectra in Orion A also found the lack of the blueshifted  tail extending beyond $v_{lsr}<0$ km s$^{-1}$.

However, while not quite as pronounced as has been reported by T09, we do observe some component of a blue-shifted tail in the stellar radial velocities relative to the $^{13}$CO motions. To determine whether or not the sources that populate that tail are members of the cluster or unrelated foreground or background stars, we searched for Li I detection towards some of them (Table \ref{tab:li}, Figure \ref{fig:co}) as a signature of their youth to establish whether or not these sources could be causally related to the ONC. In large, many sources that occupy the same velocity space as the gas and blue shifted tail have indeed been found to contain Li I. While there is some contamination from sources that appear to be somewhat more evolved, their low numbers alone cannot account for the entirety of the blue-shifted tail. On the other hand, nearly all sources that occupy velocity space outside what is presented in Figure \ref{fig:co} lack in Li I. 

There are several possible explanations for this tail. There could be a separate foreground population of young stars that is not an immediate part of the ONC. \citet{ngc1980-1} and \citet{ngc1980-2} argue that NGC 1980 is an example of such a foreground cluster. Unfortunately it is located at $83.7^\circ<\alpha<83.9^\circ$ and $-6.1^\circ<\delta<-5.8^\circ$, and the presence of the blueshifted population in that region is minimal and not spatially coherent. This is consistent with what has been found by \citet{2015dario}. However south of it, ONC is starting to turn into the L1641 cloud, thus it is possible that the tail in the southernmost regions can be attributed to this.

Some of these stars could have been dynamically scattered to achieve these velocities. While it is difficult to explain why there is no red-shifted population to make the velocity distribution symmetric, perhaps high extinction could prevent us from observing it. Inner ONC is where this effect would be the most pronounced. Not only does it have significantly higher stellar density than the rest of the cluster, allowing for more significant dynamical interactions between stars, but it is also more greatly affected by the extinction due to high density of gas. It is possible that this can account for some of the observed blue-shifted sources.

Alternatively, it is possible that the gas was being blown away by stellar feedback, leaving a somewhat older population of stars behind while newer stars formed. As suggested by T09, it is likely that in the northmost region, in the vicinity of NGC 1977, gas has been pushed back by irradiation from HD 37018, HD 37077 and HD 36958, which are B1V, B3V and B3V stars respectively, leaving behind a mini-cluster.


\citet{2009proszkow} and T09 postulate that instead the red-shift in the gas in the Trapezium and OMC 2/3 could be due to the gravitational infall of the OMC 2/3 filament towards the Trapezium cluster. This would not entirely explain the presence of the blue-shifted stellar population relative to the gas, and the location of these blue-shifted stars is not correlated with either being on or off the filament. More precise distances and proper motions that could be obtained in part by the ongoing \textit{Gaia} mission would be needed to confirm or deny the infall of the OMC 2/3.

\subsection{NGC 2264}\label{subsec:n2264}

\begin{figure}
\epsscale{0.9}
\plotone{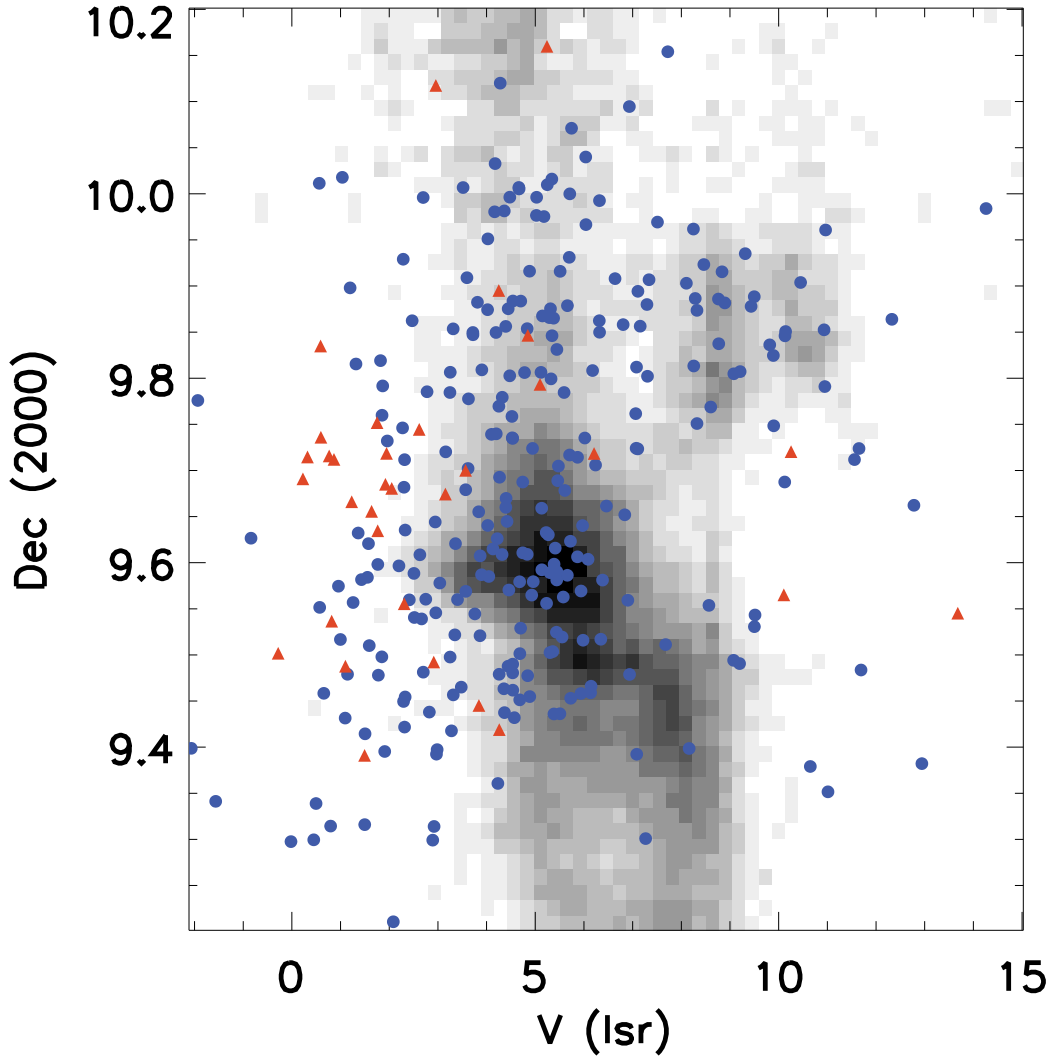} 
\caption{Position-velocity diagram for the NGC 2264 region, summed in right ascention. $^{13}$CO map from \citet{2003ridge} is plotted in the background in grayscale. All the overplotted data points are non-binary sources that were observed in at least 3 epochs. Blue dots have R.A. range between 100.05 and 100.4$^\circ$, orange triangles range between 100.4 and 100.5$^\circ$ to show a subcluster centered at $\alpha\sim100.45^\circ, \delta\sim9.7^\circ$.}\label{fig:co1}
\end{figure}

\begin{figure}
\centering
\subfigure{\includegraphics[width=0.2\textwidth]{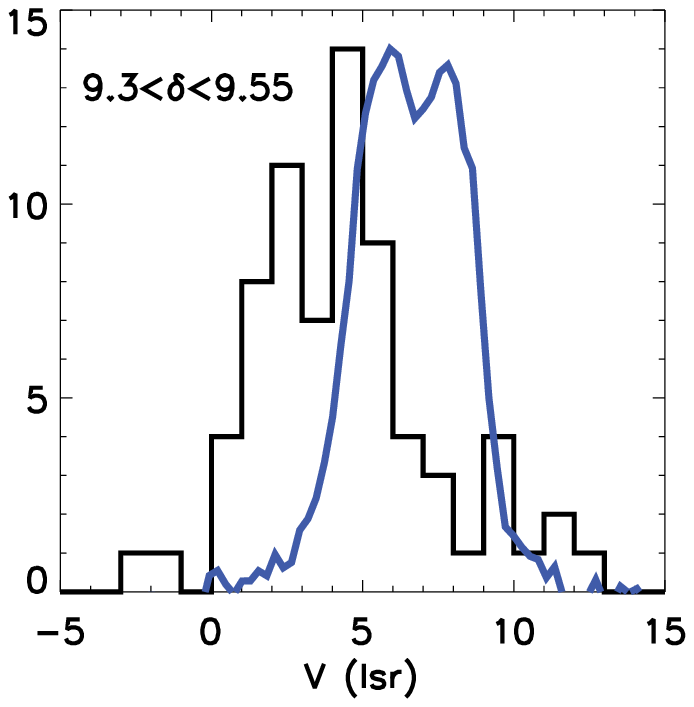}}
\subfigure{\includegraphics[width=0.2\textwidth]{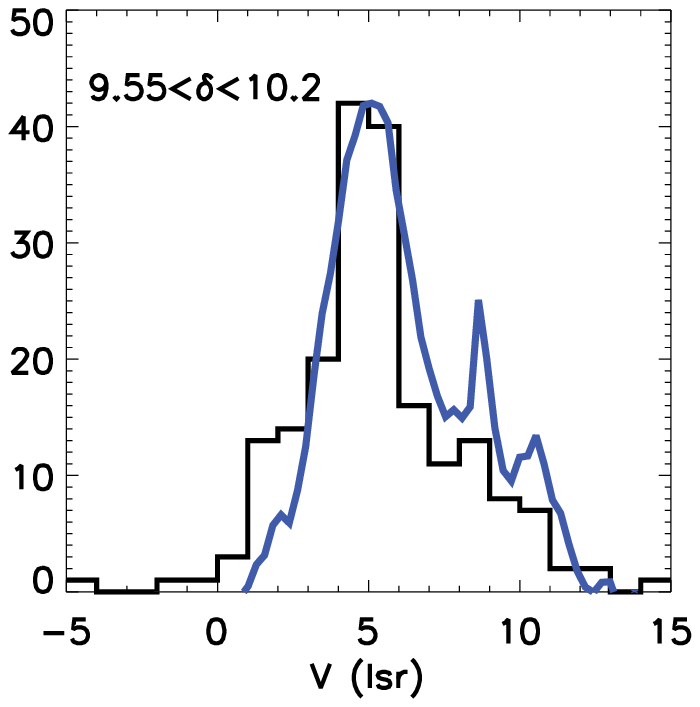}}
\caption{In black - distribution of of velocities of stars plotted in Figure \ref{fig:co1} at two declination cuts. R.A range between 100.05 and 100.4$^\circ$ has been imposed to minimize contamination from sources outside of the main cluster. In blue - summed distribution of the $^{13}$CO at those declinations, scaled to the peak of the histogram. } \label{fig:cohist1}
\end{figure}

While there is little substantial difference between velocities for NGC 2264 region quoted in T15 and this paper, some improvements could be made to previous analysis of the velocity structure for the region through better filtering of the spectroscopic binaries. Similarly as with ONC, we restrict analysis only to those sources that had been detected in at least 3 epochs and show no RV variability. Position-velocity diagram for stars is compared to that of gas from \citet{2003ridge} in Figures \ref{fig:co1} and \ref{fig:cohist1}.

We impose a constraint on sources to have R.A. of $100.05\geq\alpha\geq 100.4$ to only trace objects that are spatially correlated with $^{13}$CO gas to limit contamination from the foreground or background sources. As a result, stars that are located north of $\delta\sim9.55^\circ$ (i.e. Spokes Cluster and S Mon) have  agreement in RV with that of the gas that is significantly better than what was presented before by T15. This is partially due to stricter spatial constraints than has been originally imposed. No objects exhibit a significant blueshift in RV relative to the gas with a slight exception of the southernmost declinations in the Spokes Cluster.

However, the entirety of the stellar population found towards the Cone Nebula does show a significant blueshift that is not dissimilar to what is found towards the Trapezium and OMC 2/3 region in the ONC. Although unlike the Trapezium where the dispersion velocity of the stars is wide enough to also correlate with the gas, stars towards the Cone Nebula appear to be  decoupled from the gas. The reason for this is not entirely clear.

In addition to these regions, there appears to be a small cluster of stars centered at $\alpha\sim100.45^\circ, \delta\sim9.7^\circ$ with the diameter of $\sim0.1^\circ$ (Figure \ref{fig:co1}). It was previously identified but not discussed in T15. The members of this cluster appear to a systematic RV$_{lsr}\sim2$ km s$^{-1}$, which is somewhat distinguishable from the main cluster. It is possible that it is a either an older cluster that has managed to clear away all of its gas.

Since this cluster does not appear to be dynamically relaxed as it exhibits a significant distinct substructure, better determination of its age and further modeling will be needed to determine the degree of the interactions between these subclusters. This could shed light on the dominant method of the cluster formation, such as whether it is undergoing a cold collapse or not \citep{2002scally}.

\section{Conclusions}\label{sec:conclusion}

In this paper we continue the efforts started in T09 and T15 in characterizing stellar radial velocities of the two closest massive star-forming regions, the ONC and NGC 2264. Using multi-epoch observations we search for sources that exhibit a significant change in radial velocities that could be attributed due to a presence of a binary. We identify a total of 130 multiple system between two regions. For 8 of the sources located in NGC 2264 we produce detailed orbital fits, and for two of these sources we can determine a mass ratio between the primary and the secondary.

The multiplicity fraction that we observe is $5.8\pm 1.1$\% for the ONC, and $5.3\pm 1.2$\% for NGC 2264. If these systems were consistent with what is observed in the nearby G-dwarfs then considering uncertainties of individual measurements and allowing primaries of 0.7 M$_\odot$, we would expect to observe a MF of $4.5\pm0.9$\% and $5.8\pm1.2$\% respectively for these two clusters. Both NGC 2264 and ONC have a distribution of the multiple stars that is largely consistent with what is observed in the field within observed in the field in the same parameter space. However, a study of the wide binaries in NGC 2264 would be needed to conclusively compare the MF of these two regions.

In addition to analyzing multiplicity, we reexamined the stellar RV distribution relative to that of gas for both of these clusters to find a significantly better agreement between the two than has been previously reported, as both the peak of the distribution and velocity dispersion of stars and gas are extremely similar in many regions of these clusters. The presence of the blue-shifted young stars is reduced significantly in the cleaned sample, but they are not entirely absent. Some of these sources could be explained by the specifics of star formation processes in these regions, such as by stellar feedback pushing the gas away or by a presence of a separate foreground cluster.

\acknowledgments
J.J.T. is currently supported by grant 639.041.439 from the Netherlands Organization for Scientific Research (NWO).
\bibliographystyle{apj.bst}

\LongTables

\begin{deluxetable*}{cccccc}
\tabletypesize{\scriptsize}
\tablewidth{0pt}
\tablecaption{Dates and configurations of the observations.\tablenotemark{a}\label{tab:inst}}
\tablehead{
\colhead{Field} &\colhead{Date} & \colhead{R.A.} & \colhead{Dec.} & \colhead{Exposure time}& \colhead{Instrument}\\
\colhead{ID} &\colhead{(UT)} & \colhead{(J2000)} & \colhead{(J2000)} & \colhead{(\#$\times$seconds)}& \colhead{}
}
\startdata
F1-E1-2008 & 2008/10/19 & 05:35:23.02 & $-$04:46:26.37 & 3$\times$1200& Hectochelle \\
F1-E2-2008 & 2008/10/21 & 05:35:23.02 & $-$04:46:26.37 & 3$\times$1200& Hectochelle \\
F2-E1-2008 & 2008/10/20 & 05:35:15.14 & $-$05:15:08.42 & 3$\times$1200& Hectochelle \\
F3-E1-2008 & 2008/10/19 & 05:35:13.17 & $-$05:31:44.51 & 3$\times$1200& Hectochelle \\
F3-E2-2008 & 2008/10/21 & 05:35:13.17 & $-$05:31:44.51 & 3$\times$1200& Hectochelle \\
F4-E1-2008 & 2008/10/18 & 05:35:07.48 & $-$05:17:32.75 & 3$\times$1200& Hectochelle \\
F4-E2-2008 & 2008/10/20 & 05:35:07.48 & $-$05:17:32.75 & 3$\times$1200& Hectochelle \\
F5-E1-2008 & 2008/10/18 & 05:35:22.22 & $-$06:07:13.73 & 3$\times$1200& Hectochelle \\
F5-E2-2008 & 2008/10/20 & 05:35:22.22 & $-$06:07:13.73 & 3$\times$1200& Hectochelle \\
F1-E1-2009 & 2009/02/14 & 05:35:09.15 & $-$05:20:42.98 & 3$\times$1200& Hectochelle \\
F1-E2-2009 & 2009/11/03 & 05:35:06.94 & $-$05:17:36.21 & 3$\times$1200& Hectochelle \\
F1-E3-2009 & 2009/12/01 & 05:35:06.94 & $-$05:17:36.21 & 3$\times$1200& Hectochelle \\
F1-E4-2009 & 2009/12/03 & 05:35:06.94 & $-$05:17:36.21 & 3$\times$1200& Hectochelle \\
F2-E1-2009 & 2009/03/14 & 05:35:14.69 & $-$05:04:58.26 & 3$\times$1200& Hectochelle \\
F3-E1-2009 & 2009/12/02 & 05:34:52.35 & $-$05:54:23.11 & 3$\times$1200& Hectochelle \\
F4-E1-2009 & 2009/12/02 & 05:35:26.82 & $-$06:09:44.54 & 3$\times$1200& Hectochelle \\
F5-E1-2009 & 2009/12/02 & 05:35:09.76 & $-$05:16:54.04 & 3$\times$1200& Hectochelle \\
F6-E1-2009 & 2009/12/02 & 05:35:20.82 & $-$04:49:07.77 & 3$\times$1200& Hectochelle \\ 
OA & 2008/11/06 & 05:35:07.2 & $-$05:52:14.2 & 4$\times$1200 & MIKE \\
OB & 2008/11/07 & 05:35:00.0 & $-$05:25:18.4 & 4$\times$1200 & MIKE \\
OC & 2008/11/07 & 05:35:26.9 & $-$05:13:13.2 & 4$\times$1200 & MIKE \\
OD & 2008/11/06 & 05:35:26.9 & $-$04:47:34.7 & 5$\times$1200 & MIKE \\
OA1 & 2014/02/21 & 5:35:12.00 & $-$5:30:00.0 & 6$\times$600 &  M2FS (Mg)\\
OB1 & 2013/12/01 & 5:35:24.61 & $-$5:11:58.2 & 4$\times$600 &  M2FS (Mg)\\
OC1 & 2013/11/26 & 5:35:12.00 & $-$6:00:00.0 & 4$\times$600 &  M2FS (Mg)\\
OD1 & 2013/11/26 & 5:35:24.00 & $-$4:45:00.0 & 5$\times$600 &  M2FS (Mg)\\
NA & 2014/02/23 & 6:40:25.48 & +9:48:26.0 & 5$\times$600 &  M2FS (Mg)\\
NB & 2014/02/25 & 6:41:19.45 & +9:30:28.6 & 5$\times$600 & M2FS (Mg) \\
LOA & 2014/12/18 & 5:35:12.00 & $-$5:18:04.0 & 6$\times$600 &  M2FS (Li)\\
LOB & 2014/12/21 & 5:35:09.00 & $-$6:02:00.6 & 3$\times$1200 &  M2FS (Li)\\
LOC & 2014/12/24 & 5:35:22.90 & $-$4:43:27.8 & 4$\times$1200 &  M2FS (Li)\\
\enddata
\tablenotetext{a}{Data that have been presented in T09 and T15 is not listed in this table.}
\end{deluxetable*}
\begin{deluxetable*}{cccccccccc}
\tabletypesize{\scriptsize}
\tablewidth{0pt}
\tablecaption{Sources that show no variation in radial velocity between multiple observations.\tablenotemark{a}\label{tab:single}}
\tablehead{
\colhead{RV}& \colhead{R.A.} &\colhead{Dec.} & \colhead{Date} & \colhead{$v$} & \colhead{$\sigma$} & \colhead{$R$}& \colhead{Temp}& \colhead{RR?\tablenotemark{b}}& \colhead{Instrument}\\
\colhead{$\#$}&\colhead{(J2000)} &\colhead{(J2000)} & \colhead{(JD)} & \colhead{(km s$^{-1}$)} & \colhead{km s$^{-1}$}  & \colhead{}&  \colhead{(K)}& \colhead{}& \colhead{}
}
\startdata
1 & 05:33:17.95 & -05:21:38.6 & 2454401.0 &  29.17 &  4.51 &   3.26 & 3500 & --- & Hectochelle \\
 & &  & 2454401.8 &  30.38 &  4.55 &   3.07 & 3750 & --- & Hectochelle \\
 & &  & 2454757.9 &  23.64 &  3.14 &   6.05 & 3750 & --- & Hectochelle \\
 & &  & 2454760.0 &  25.86 &  3.88 &   5.63 & 3750 & --- & Hectochelle \\
 & &  & 2454876.6 &  24.78 &  2.58 &   7.42 & 3750 & --- & Hectochelle \\
\hline
2 & 05:33:20.44 & -05:11:24.0 & 2454757.9 &  25.83 &  1.07 &  12.19 & 4000 & --- & Hectochelle \\
 & &  & 2454760.0 &  26.79 &  1.18 &  13.83 & 4000 & y & Hectochelle \\
 & &  & 2454904.6 &  24.57 &  1.12 &  14.51 & 4000 & y & Hectochelle \\
 & &  & 2455167.9 &  31.86 &  1.43 &   8.73 & 4000 & y & Hectochelle \\
\hline
3 & 05:33:22.58 & -05:32:40.1 & 2454402.9 &  23.16 &  1.51 &   7.74 & 4000 & --- & Hectochelle \\
 & &  & 2454876.6 &  23.32 &  0.78 &  17.49 & 4000 & --- & Hectochelle \\
\enddata
\tablenotetext{a}{Includes sources with insufficient number of detections to determine variability.}
\tablenotetext{b}{Measurement was processed with low\_bin=3, top\_low=10}
\tablenotetext{c}{RV 1-1154 belong to the ONC, RV 1155-2057 to NGC 2264.}
\tablenotetext{}{Full version of the table will be available in the online text.}
\end{deluxetable*}
\clearpage
\begin{deluxetable*}{cccccccccccc}
\tabletypesize{\scriptsize}
\tablewidth{0pt}
\tablecaption{Sources with variable radial velocity.\label{tab:binary}}
\tablehead{
\colhead{RV}& \colhead{R.A.} &\colhead{Dec.} & \colhead{Date} & \colhead{$v_1$} & \colhead{$\sigma_1$} & \colhead{$v_2$\tablenotemark{a}} & \colhead{$\sigma_2$} & \colhead{$R$}& \colhead{Temp}& \colhead{RR?\tablenotemark{b}}& \colhead{Instrument}\\
\colhead{$\#$}&\colhead{(J2000)} &\colhead{(J2000)} & \colhead{(JD)} & \colhead{(km s$^{-1}$)} & \colhead{km s$^{-1}$} & \colhead{(km s$^{-1}$)} & \colhead{km s$^{-1}$} & \colhead{}&  \colhead{(K)}& \colhead{}& \colhead{}
}
\startdata
10\tablenotemark{c} & 05:33:29.38 & -05:07:49.1 & 2454401.0 &  31.39 &  1.95 & --- & --- &   8.24 & 5000 & --- & Hectochelle \\
 & &  & 2454401.8 &  29.55 &  1.66 & --- & --- &  10.62 & 4750 & --- & Hectochelle \\
 & &  & 2454757.9 &  30.34 &  1.44 & --- & --- &  11.24 & 5000 & --- & Hectochelle \\
 & &  & 2454760.0 &  29.20 &  1.31 & --- & --- &  12.47 & 4250 & --- & Hectochelle \\
 & &  & 2455138.8 &  19.19 &  1.01 & --- & --- &  13.12 & 5000 & --- & Hectochelle \\
 & &  & 2455168.9 &  32.64 &  1.46 & --- & --- &  10.28 & 4250 & --- & Hectochelle \\
\hline
26 & 05:33:36.37 & -05:01:40.5 & 2454759.9 &  99.50 &  2.40 & --- & --- &  12.37 & 5750 & --- & Hectochelle \\
 & &  & 2455138.8 &  20.43 &  1.32 & --- & --- &  10.17 & 5000 & --- & Hectochelle \\
 & &  & 2455166.9 &  72.24 &  6.11 & --- & --- &   4.43 & 5500 & --- & Hectochelle \\
\hline
41 & 05:33:41.88 & -05:08:17.1 & 2454757.9 &  32.51 &  0.76 & --- & --- &  16.07 & 6750 & y & Hectochelle \\
 & &  & 2454760.0 &  33.27 &  0.92 & --- & --- &  13.31 & 6500 & y & Hectochelle \\
 & &  & 2455138.8 &  22.54 &  0.96 & --- & --- &  13.22 & 5500 & --- & Hectochelle \\
 & &  & 2455166.9 &  30.96 &  1.87 & --- & --- &   6.43 & 7000 & --- & Hectochelle \\
 & &  & 2455168.9 &  32.77 &  1.02 & --- & --- &  12.34 & 6750 & --- & Hectochelle \\
\enddata
\tablenotetext{a}{Velocity obtained from the second peak of the cross-correlation for double-lined binaries.}
\tablenotetext{b}{Measurement was processed with low\_bin=3, top\_low=10}
\tablenotetext{c}{Have at least 4 $R>6$ measurements separated by more than 3 days with only single variable velocity.}
\tablenotetext{}{Full version of the table will be available in the online text.}
\end{deluxetable*}
\vspace{10 mm}
\begin{deluxetable*}{ccccccccc}
\tabletypesize{\scriptsize}
\tablewidth{0pt}
\tablecaption{Sources that were surveyed for the presence of Li I. \label{tab:li}}
\tablehead{
\colhead{RV} &\colhead{R.A.} &\colhead{Dec.} & \colhead{$v_{Li}$\tablenotemark{b}} & \colhead{$\sigma_{Li}$} & \colhead{$R_{Li}$} & \colhead{$\lambda_{Li}$\tablenotemark{c}}& \colhead{$W_{Li}$\tablenotemark{d}} & \colhead{Temp}\\
\colhead{$\#$\tablenotemark{a}} &\colhead{(J2000)} &\colhead{(J2000)} & \colhead{(km s$^{-1}$)} & \colhead{(km s$^{-1}$)} & \colhead{} & \colhead{\AA}& \colhead{\AA}& \colhead{K.}
}
\startdata
113 & 05:34:15.45 & -06:06:55.1 &  49.51 &  0.72 &  10.58 & --- & --- & 3750 \\
120 & 05:34:17.78 & -05:55:43.1 &   4.52 &  0.99 &   7.69 & --- & --- & 6000 \\
133 & 05:34:20.80 & -05:23:29.2 &  18.62 &  1.48 &   4.87 & 6708.32 & 0.39 & 3500 \\
160 & 05:34:27.34 & -05:24:22.2 &  26.02 &  2.26 &   9.18 & 6708.46 & 0.47 & 4000 \\
164 & 05:34:28.22 & -05:59:09.0 &  33.56 &  0.82 &  10.78 & 6708.62 & 0.10 & 6250 \\
187 & 05:34:33.01 & -05:57:47.1 &  26.96 &  3.82 &   6.73 & 6708.53 & 0.53 & 4000 \\
192 & 05:34:33.86 & -05:56:38.0 &  24.43 &  0.64 &  11.71 & 6708.48 & 0.62 & 3500 \\
200 & 05:34:35.16 & -05:58:15.3 &  51.69 &  0.52 &  15.64 & --- & --- & 5250 \\
\enddata
\tablenotetext{a}{Sources with RV$\#$ greater than 2057 have $R<6$ for all detections, and thus they are not included in the Tables \ref{tab:single} and \ref{tab:binary}}
\tablenotetext{b}{Velocity and other properties were measured only from Li I data.}
\tablenotetext{c}{Typical uncertainty in $\lambda_{Li}$ is 0.01 \AA}
\tablenotetext{d}{Typical uncertainty in $W_{Li}$ is 0.01 \AA}
\tablenotetext{}{Full version of the table will be available in the online text.}
\end{deluxetable*}

\clearpage

\clearpage
\begin{deluxetable}{lrclrcl}
\tabletypesize{\scriptsize}
\tablewidth{0pt}
\tablecaption{Orbital parameters for double-lined binaries\label{tab:dbpeak}}
\tablehead{
\colhead{Parameter} &\multicolumn{3}{c}{RV 1768} &\multicolumn{3}{c}{RV 1659}
}
\startdata
\multicolumn{7}{c}{Adjusted Quantities}\\
\hline
$P$ (d)		&4.7878 &$\pm$& 0.0002		&15.3182 &$\pm$& 0.0007\\
$T_p$ (HJD)		&2454289.50 &$\pm$& 0.06		&2454294.07 &$\pm$& 0.05\\
$e$			&0.55 &$\pm$& 0.05			&0.00 &$\pm$& 0.02\\
$\omega$ (deg)		&139.68 &$\pm$& 3.16		&0.00 &$\pm$& 0.60\\
$\gamma$ (km/s)	&16.05 &$\pm$& 0.54	&22.83 &$\pm$& 0.46\\
$K_1$ (km/s)		&59.34 &$\pm$& 6.34		&41.56 &$\pm$& 1.07\\
$K_2$ (km/s)		&60.83 &$\pm$& 6.32		&52.20 &$\pm$& 1.05\\
\hline
\multicolumn{7}{c}{Derived Quantities}\\
\hline
$M_1\sin ^3i$ ($M_\odot$)	&0.254 &$\pm$& 0.066	&0.73 &$\pm$& 0.04\\
$M_2\sin ^3i$ ($M_\odot$)	&0.248 &$\pm$& 0.065	&0.58 &$\pm$& 0.03\\
$q = M_2/M_1$		&0.98 &$\pm$& 0.15		&0.80 &$\pm$& 0.03\\
$a_1\sin i$ ($10^6$ km)	&3.26 &$\pm$& 0.37	&8.75 &$\pm$& 0.23\\
$a_2\sin i$ ($10^6$ km)	&3.35 &$\pm$& 0.37	&11.00 &$\pm$& 0.22\\
$a  \sin i$ ($10^6$ km)	&6.61 &$\pm$& 0.52	&19.75 &$\pm$& 0.32\\
\hline
\multicolumn{7}{c}{Other Quantities}\\
\hline
$\chi^2$			&\multicolumn{3}{c}{8.38}			&\multicolumn{3}{c}{13.43}\\
$N_{obs}$ (primary)		&\multicolumn{3}{c}{10}			&\multicolumn{3}{c}{9}\\
$N_{obs}$ (secondary)	&\multicolumn{3}{c}{8}			&\multicolumn{3}{c}{8}\\
Time span (days)		&\multicolumn{3}{c}{2308.7}			&\multicolumn{3}{c}{2308.6}\\
$rms_1$ (km/s)		&\multicolumn{3}{c}{1.62}			&\multicolumn{3}{c}{1.22}\\
$rms_2$ (km/s)		&\multicolumn{3}{c}{1.42}			&\multicolumn{3}{c}{2.10}\\
T$_{ave}$ (K) &\multicolumn{3}{c}{4021} &\multicolumn{3}{c}{4203}
\enddata
\end{deluxetable}
\clearpage

\newpage


\clearpage
\begin{tiny}

\clearpage
\begin{turnpage}
\begin{deluxetable*}{lrclrclrclrclrclrcl}
\tabletypesize{\tiny}
\tablewidth{0pt}
\tablecaption{Orbital parameters for single-lined binaries\label{tab:speak}}
\tablehead{
\colhead{Parameter} &\multicolumn{3}{c}{RV 1166} &\multicolumn{3}{c}{RV 1372} &\multicolumn{3}{c}{RV 1496} &\multicolumn{3}{c}{RV 1550} &\multicolumn{3}{c}{RV 1660} &\multicolumn{3}{c}{RV 1753}
}
\startdata
\multicolumn{19}{c}{Adjusted Quantities}\\
\hline
$P$ (d)		&105.82 &$\pm$& 0.27		&315.41 &$\pm$& 5.79		&588.93 &$\pm$& 9.47		&72.83 &$\pm$& 0.03		&622.00 &$\pm$& 26.32		&12.93 &$\pm$& 0.01\\
$T_p$ (HJD)		&2452620.5 &$\pm$& 29.3		&2454382.4 &$\pm$& 5.5		&2454814.5 &$\pm$& 22.1		&2454329.0 &$\pm$& 1.1		&2454585.1 &$\pm$& 31.4		&2454299.3 &$\pm$& 0.6\\
$e$			&0.00 &$\pm$& 0.07			&0.65 &$\pm$& 0.12			&0.44 &$\pm$& 0.07			&0.32 &$\pm$& 0.02			&0.56 &$\pm$& 0.08			&0.170 &$\pm$& 0.03\\
$\omega$ (deg)		&224.02 &$\pm$& 99.11		&178.80 &$\pm$& 5.31		&8.61 &$\pm$& 11.29		&266.43 &$\pm$& 7.12		&135.59 &$\pm$& 11.99		&194.66 &$\pm$& 19.74\\
$\gamma$ (km/s)	&15.93 &$\pm$& 0.25	&-29.97 &$\pm$& 0.78 &28.05 &$\pm$& 0.30	&20.80 &$\pm$& 0.26	&23.08 &$\pm$& 0.39	&17.97 &$\pm$& 1.35\\
$K_1$ (km/s)		&4.24 &$\pm$& 0.46		&20.00 &$\pm$& 7.73		&3.46 &$\pm$& 0.87		&16.91 &$\pm$& 0.39		&5.86 &$\pm$& 0.51		&16.11 &$\pm$& 1.61\\
\hline
\multicolumn{19}{c}{Derived Quantities}\\
\hline
$a_1\sin i$ ($10^6$ km)	&6.17 &$\pm$& 0.67	&65.91 &$\pm$& 27.09	&25.09 &$\pm$& 6.38	&16.06 &$\pm$& 0.39 	&41.36 &$\pm$& 4.93	&2.82 &$\pm$& 0.28\\
$f(m_1,m_2)$ ($M_\odot$)	&0.0008 &$\pm$& 0.0003	&0.11 &$\pm$& 0.15	&0.0018 &$\pm$& 0.0014	&0.031 &$\pm$& 0.003	&0.007 &$\pm$& 0.003	&0.005 &$\pm$& 0.002\\
\hline
\multicolumn{19}{c}{Other Quantities}\\
\hline
$\chi^2$		&\multicolumn{3}{c}{1.31}		&\multicolumn{3}{c}{2.61}		&\multicolumn{3}{c}{1.74}		&\multicolumn{3}{c}{1.76}		&\multicolumn{3}{c}{2.61}		&\multicolumn{3}{c}{2.61}\\
$N_{obs}$ (primary)	&\multicolumn{3}{c}{10}		&\multicolumn{3}{c}{11}		&\multicolumn{3}{c}{10}		&\multicolumn{3}{c}{11}		&\multicolumn{3}{c}{11}		&\multicolumn{3}{c}{11}\\
Time span (days)	&\multicolumn{3}{c}{903.6}		&\multicolumn{3}{c}{2308.7}		&\multicolumn{3}{c}{2327.7}		&\multicolumn{3}{c}{2325.7}		&\multicolumn{3}{c}{2325.6}		&\multicolumn{3}{c}{2327.7}\\
$rms_1$ (km/s)	&\multicolumn{3}{c}{0.17}		&\multicolumn{3}{c}{0.53}		&\multicolumn{3}{c}{0.19}		&\multicolumn{3}{c}{0.25}		&\multicolumn{3}{c}{0.55}		&\multicolumn{3}{c}{0.28}\\
T$_{ave}$ (K) &\multicolumn{3}{c}{6274} &\multicolumn{3}{c}{4267} &\multicolumn{3}{c}{5577} &\multicolumn{3}{c}{4361} &\multicolumn{3}{c}{4256} &\multicolumn{3}{c}{5301}
\enddata
\end{deluxetable*}
\end{turnpage}
\clearpage

\end{tiny}
\end{document}